# The FENIKS Survey: Stellar–Halo Mass Relationship of Central and Satellite Galaxies in UDS and COSMOS at $0.2 < z < 4.5$

Kumail Zaidi[1,2], David A. Wake[3,4], Danilo Marchesini[1], Kartheik Iyer[5,6], Adam Muzzin[7], Casey Papovich[8,9], Jacqueline Antwi-Danso[5], Karl Glazebrook[10,11], and Ivo Labbé[11]

[1] Tufts University, 574 Boston Avenue, Medford, MA 02155, USA; kumail.tufts@gmail.com
[2] HEP Division, Argonne National Laboratory, 9700 South Cass Avenue, Lemont, IL 60439, USA
[3] Department of Physics and Astronomy, University of North Carolina Asheville, Asheville, NC 28804, USA
[4] Department of Physics and Astronomy, University College London, Gower St., London WC1E 6BT, UK
[5] David A. Dunlap Department of Astronomy & Astrophysics, University of Toronto, 50 St George St, Toronto, ON M5S 3H4, Canada
[6] Columbia Astrophysics Laboratory, Columbia University, 550 West 120th Street, New York, NY 10027, USA
[7] Department of Physics & Astronomy, York University, 4700 Keele Street Toronto, Ontario, M3J 1P3, Canada
[8] George P. and Cynthia Woods Mitchell Institute for Fundamental Physics and Astronomy, Texas A&M University, College Station, TX 78743, USA
[9] Department of Physics & Astronomy, Texas A&M University, 4242 TAMU, College Station, TX 78743, USA
[10] ARC Centre of Excellence for All Sky Astrophysics in 3 Dimensions (ASTRO 3D), Australia
[11] Centre for Astrophysics and Supercomputing, Swinburne University of Technology, Melbourne, VIC 3122, Australia
*Received 2024 November 6; revised 2026 March 28; accepted 2026 May 7; published 2026 June 8*

## Abstract

We present the observed stellar–halo mass relationship (SHMR) spanning $0.2 < z < 4.5$ enabled through galaxy clustering and abundance measurements from two large (effective area $\sim$1.61 deg$^2$) and homogeneously prepared photometric catalogs—UltraVISTA ultradeep stripes DR3 (COSMOS) and FENIKS v2 (UDS). To translate these measurements into the SHMR, we introduce a novel halo occupation distribution (HOD) fitting approach ("smooth-$z$") whereby HOD parameters between neighboring $z$-bins are disallowed from sharp changes through continuity priors. As a result, the high constraining power at $z \lesssim 2$, due to a much wider dynamical range in stellar mass ($\sim$3 dex), helps constrain the SHMR at $z \gtrsim 2$, where that range shrinks down to $\lesssim$1 dex. We find a clear downsizing trend in halo mass scale with peak integrated star-forming efficiency (SFE), $M_h^{\rm peak}$, as it consistently decreases from $10^{12.69^{+0.08}_{-0.08}}\ M_\odot$at $z \sim 4$ to $10^{12.17^{+0.03}_{-0.03}}\ M_\odot$at $z \sim 0.35$, exhibiting a decline of 0.52 dex. Furthermore, we show that if we had relied on COSMOS data alone (as opposed to UDS+COSMOS), as has been done by many preceding studies, $M_h^{\rm peak}$ would be systematically lower by up to $\sim$0.2 dex at $z < 1.25$, highlighting the importance of mitigating cosmic variance. Finally, for the first time, we show how the integrated SFE evolves with redshift as halos grow in mass along their progenitor merger trees instead of at fixed halo masses.

## 1. Introduction

In the standard cold dark matter with dark energy ($\Lambda$CDM) model, the structure of the Universe forms hierarchically. Small seeds of primordial overdensities in the structure of the Universe, as evidenced through the cosmic microwave background radiation (CMB), grow larger over time. Since dark matter accounts for $\sim$84% of the matter in the Universe (Planck Collaboration et al. 2020), through gravity, it guides the bulk of this evolution. Regions overdense in dark matter grow in size, accumulate enough mass to become self-gravitating halos, and decouple from the expansion of the Universe caused by dark energy. Within the potential wells of these halos, cold streams of gas funnel in, forming cold and dense clumps eventually reaching pressure, temperature, and density thresholds to trigger nuclear fusion, star formation, and galaxy formation (S. D. M. White & M. J. Rees 1978; S. M. Fall & G. Efstathiou 1980; V. Springel et al. 2006).

Over time, halos grow in size and mass through merging and so do their resident galaxies. The evolution of dark matter halos and galaxies are intricately linked in this manner. To understand the influence of the aforementioned structure formation on galaxy formation and vice versa, we need to understand how galaxy properties like stellar mass are dictated by their host dark matter halo properties like mass, the so-called galaxy–halo connection (see R. H. Wechsler & J. L. Tinker (2018) for a review), and vice versa. Our first hint into the galaxy–halo connection comes by comparing the abundance of halos—halo mass functions (HMFs)—from $N$-body simulations with the observed abundance of galaxies—galaxy stellar mass functions (SMFs; see, e.g., L. Legrand et al. (2019) for a relatively recent example). Through this approach, it was realized early that galaxy formation is indeed a very inefficient process (e.g., C. Conroy & R. H. Wechsler 2009). The SMFs have a much flatter slope at low mass than the HMFs, indicating inefficient star formation, and on the other hand, a much sharper decline at higher masses than the HMFs, again indicating inefficient star formation.

Another facet of this inefficient star formation is the observationally constrained stellar–halo mass relationship (SHMR) itself (R. Mandelbaum et al. 2006; B. P. Moster et al. 2010). The resulting stellar–halo mass ratio ($M_*/M_h$) at a given epoch is a strong indicator of how efficient the halos with the current mass, $M_h$, have been throughout their history in converting available baryons into stars ("in situ" star formation), and in the accretion of stellar matter through mergers ("ex situ" star formation). $M_*/M_h$ can be easily converted into this star-forming efficiency (SFE) by dividing it with the available baryonic fraction of the Universe, $f_b = \Omega_b/\Omega_m = 0.158$ (Planck Collaboration et al. 2020), where $\Omega_b$ and $\Omega_m$ are the baryonic and total matter (baryons + dark matter) density fractions.

Numerous low- and high-redshift observational data-based studies (A. Leauthaud et al. 2012 out to $z \sim 1$; J. Coupon et al. 2012 out to $z \sim 1.2$; D. A. Wake et al. 2011 at $1 < z < 2$; H. J. McCracken et al. 2015 out to $z \sim 2$; P. W. Hatfield et al. 2016 at $0.5 < z < 1.7$; W. I. Cowley et al. 2018 at $1.5 \lesssim z \lesssim 3$; L. Legrand et al. 2019 out to $z \sim 5$; M. Shuntov et al. 2022 out to $z \sim 5$; P. S. Behroozi et al. 2013 out to $z \sim 8$; and P. Behroozi et al. 2019 out to $z \sim 10$) either directly show or support that SFE remains low at all redshifts—on the order of a few percent. It peaks at most at $\sim$20 % at $M_h \sim 10^{12}\ M_\odot$, a mass which evolves mildly with redshift. At lower and higher halo masses, SFE drops sharply due to various feedback processes that suppress star formation. At the low-mass end, it is thought to be stellar winds that expel or heat the surrounding gas and hot bubbles created by supernovae explosions that limit star formation (P. F. Hopkins et al. 2012). At the high-mass end, active galactic nuclei (AGN) feedback specifically in "radio mode" (radiatively inefficient, low accretion rate) is thought to shut down star formation (J. Silk & M. J. Rees 1998; D. J. Croton et al. 2006; R. Kondapally et al. 2023; L. Scharré et al. 2024). This suppression in star formation at lower and higher halo masses relative to $\sim 10^{12}\ M_\odot$ also becomes visible in the deviation of the slopes at low/high mass of observed SMFs from the cosmological simulation-derived HMFs, as mentioned above.

As the observationally constrained SHMR is a result of the amalgamation of physical processes at various mass scales, it provides strong priors for the cutting-edge semianalytic models such as Dark SAGE (A. R. H. Stevens et al. 2016, 2024) and L-GALAXIES (B. M. B. Henriques et al. 2015, 2020), as well as suites of state-of-the-art hydrodynamical simulations such as IllustrisTNG (R. Weinberger et al. 2017; A. Pillepich et al. 2018) and the FIRE project (A. Wetzel et al. 2023; P. F. Hopkins et al. 2023). This unison of observations and simulations has been incredibly helpful in advancing our understanding of the galaxy–halo connection (R. S. Somerville & J. R. Primack 1999; S. Cole et al. 2000; V. Springel et al. 2005; R. G. Bower et al. 2006; D. J. Croton et al. 2006; G. De Lucia & J. Blaizot 2007; Q. Guo et al. 2011; M. Vogelsberger et al. 2014; J. Schaye et al. 2015; A. Pillepich et al. 2018). While the hydrodynamical simulations and semianalytical models are incredibly helpful in informing the baseline understanding of the galaxy–halo connection, their conclusions are heavily dependent on assumptions related to gas dynamics, treatment of supernovae, and the implementation of the chosen subgrid recipes beyond the resolution scale of the simulation, among others.

A more observation-dependent approach to constraining the relationship between assembly histories of dark matter halos and their resident galaxies is the so-called empirical modeling where physical priors come directly from observations. Several observational probes of galaxy formation such as star formation rates, SMFs, quenched fractions, galaxy clustering, etc., are utilized in a self-consistent way to inform the creation of mock universes by populating halo merger trees (with galaxies) from dark-matter-only simulations (P. S. Behroozi et al. 2013; B. P. Moster et al. 2018; P. Behroozi et al. 2019). Even though empirical models do not assume any models of galaxy formation, they adopt parameterizations to capture, e.g., star formation in halos based on halo properties and then further parameterize their redshift evolution; they too remain quite model dependent. Moreover, the interobservational systematics, resulting from combining large amounts of data from several surveys with differently calibrated data reduction pipelines, are bound to skew the results.

Perhaps the simplest approach to connecting galaxies to their host halos is to match their abundances—"abundance matching" (C. Marinoni & M. J. Hudson 2002; C. Conroy & R. H. Wechsler 2009; P. S. Behroozi et al. 2010). In this method, the halos are usually rank ordered by their mass and assigned the rank-ordered galaxies by stellar mass (or luminosity) individually. The idea is simple: the more massive a galaxy, the more massive its host. However, just matching abundances ignores the galaxy–halo connection information hidden in the structure on all scales which can be measured via clustering. Clustering on small scales contains information on how the satellite galaxies are distributed within a halo and on large scales contains information of how both the central and satellites are distributed. Furthermore, since clustering is directly related to halo mass, it is an indirect probe of the halo mass, which, combined with the information contained in SMFs, provides tighter constraints on the galaxy–halo connection (P. S. Behroozi et al. 2010; J. Coupon et al. 2015). Finally, if we keep the abundances of stellar mass and halo mass unchanged but tweak the scatter in stellar mass at fixed halo mass, we will assign the galaxies to halos differently resulting in a change in clustering (R. H. Wechsler & J. L. Tinker 2018). Therefore, measurements of clustering, beyond helping constrain the halo mass, allow us to constrain the scatter in stellar mass (at fixed halo mass) better.

The observations of both galaxy clustering and abundance can be leveraged to break degeneracies as mentioned above to get a robust estimate of SHMR within the framework of halo occupation distribution (HOD) modeling. The HOD model simply populates halos of different masses with different numbers of galaxies, on average, and is split between contributions from central and satellite galaxies in the traditional HOD models (Z. Zheng et al. 2005, 2007; I. Zehavi et al. 2005, 2011; A. Leauthaud et al. 2011; Y. Zu & R. Mandelbaum 2015). Besides having centrals and satellites, galaxies can be further split into contributions from star-forming and passive galaxies as in the J. L. Tinker et al. (2013) and Y. Zu & R. Mandelbaum (2016) HOD models.

A chosen HOD model in combination with an adopted HMF from simulations and a few more ingredients (halo density profile, halo mass-concentration relation, etc.) can be used to model a power spectrum that would be produced by a distribution of galaxies on the sky, or, in other words, galaxy clustering. The HOD parameters that produce this clustering and the abundance can then be fine-tuned by fitting against the observations of galaxy clustering and the highly constraining galaxy abundance (SMFs). A constrained HOD model then gives the SHMR and many other interesting quantities such as the fraction of galaxies that are satellites as opposed to centrals in a halo of a given mass. HOD model fitted at different redshifts allows us to trace the assembly of central and satellite galaxies and, since it does not assume any galaxy formation physics but is purely observation-dependent in that regard, it gives us clues into unknown astrophysical phenomena that might be at play. This is especially true at high redshift where our understanding of overall galaxy formation remains quite extrapolated.

Besides being constrained by the observations of galaxy clustering and abundance, the HOD model can additionally incorporate constraints from galaxy–galaxy lensing observations. For instance, A. Leauthaud et al. (2012) combined all of the three probes: galaxy clustering, galaxy abundance, and galaxy–galaxy lensing to constrain the HOD, and thus SHMR out to $z \sim 1$ using data in the COSMOS field (N. Scoville et al. 2007a), although lensing observations prove difficult at $z > 1$.

To set the stage for our paper, below we highlight a few recent contributions in the literature that constrain the SHMR and the corresponding $M_h^{\rm peak}$ through observations of clustering and abundance under the framework of HOD.

S. Ishikawa et al. (2017) measured the SHMR by fitting the Z. Zheng et al. (2005) HOD with clustering and abundance of *u*-, *g*-, and *r*-dropout galaxies at $z \sim 3$, 4, and 5, respectively. They find that $M_h^{\rm peak}$ climbs up steadily from $10^{11.92 \pm 0.097} M_\odot$ at $z \sim 5$ to $10^{12.25 \pm 0.053} M_\odot$ at $z \sim 3$, opposite of the expectation from downsizing (L. L. Cowie et al. 1996), in which case the $M_h^{\rm peak}$ should decline with time. However, this was in agreement with the empirical modeling (P. S. Behroozi et al. 2013).

Using the Z. Zheng et al. (2007) HOD, W. I. Cowley et al. (2018) probed the SHMR at $1.5 \lesssim z \lesssim 5$ leveraging the Spitzer IRAC 3.6 and 4.5 $\mu$m data (SMUVS survey; M. L. N. Ashby et al. 2018) within the UltraVISTA ultradeep stripes (H. J. McCracken et al. 2012) of the COSMOS field (N. Scoville et al. 2007a). They were the first to use a stellar mass-selected sample to constrain the $M_h^{\rm peak}$ at $z \sim 2.5$, which was found to be $10^{12.5^{+0.10}_{-0.08}} M_\odot$, a mild increase from lower redshifts, consistent with downsizing.

Recently, using the A. Leauthaud et al. (2011, hereafter L11) HOD, M. Shuntov et al. (2022) constrained the SHMR out to $z \sim 5$ by utilizing a large homogeneously prepared catalog, COSMOS2020 (J. R. Weaver et al. 2022), covering an effective area of 1.27 deg$^2$ in a single field. For the first time with a stellar mass-selected sample, they extended the redshift frontier for $M_h^{\rm peak}$ out to $z \sim 5$, showing a drop of $\sim$1 dex going from $10^{12.94^{+0.36}_{-0.34}}$ $M_\odot$ at $z \sim 5$ to $10^{12.07^{+0.07}_{-0.07}} M_\odot$ at $z \sim 0.35$, a strong downsizing (L. L. Cowie et al. 1996) signal indeed. However, their uncertainty on $M_h^{\rm peak}$ at $z \sim 5$ spans a large range of $\sim$0.8 dex. This is primarily because the sample remains complete only for the massive galaxies ($M_* \gtrsim 10^{10}$ $M_\odot$) at high z, thus providing no constraints on the low-mass end of the SHMR.

In this paper, we constrain the relationship between galaxy stellar mass and their host dark matter halo mass using HOD modeling. Out to $z \sim 5$, this study is the first to use a large total effective area of $\sim$1.61 deg$^2$ evenly split between two fields: UDS $\sim$ 0.79 deg$^2$ and COSMOS $\sim$ 0.82 deg$^2$, which greatly reduces the effect of cosmic variance in measurements. Much of the previous work in observationally deriving SHMR is based on the single COSMOS field or dominated by the COSMOS field (e.g., W. I. Cowley et al. 2018; L. Legrand et al. 2019; M. Shuntov et al. 2022). The availability of deep optical to mid-IR photometry in both fields leads to robust redshifts and stellar population properties allowing accurate measurement of galaxy clustering and abundance. More importantly, the photometric catalogs in both fields are homogeneously prepared so they are free from systematic biases affecting the interpreted SHMR (P. S. Behroozi et al. 2013; B. P. Moster et al. 2018; P. Behroozi et al. 2019).

Another novel aspect of this work is to jointly fit the galaxy clustering and abundance data in the entire redshift range of $0.2 < z < 4.5$ by imposing smooth continuity priors on the redshift evolution of the HOD parameters. Tying together the *z*-bins in this way allows us to constrain the evolution of SHMR with higher confidence, especially at high redshift $z \gtrsim 2$ where the samples are only complete for the most massive galaxies ($M_* \gtrsim 10^{10}$ $M_\odot$). This would not have been possible if, following the usual procedure in the literature (e.g., M. Shuntov et al. 2022), we had solely relied on the weaker constraining power of the individual *z*-bins at high redshift. Finally, we trace the SFE of halos as they grow in mass along their merger trees in the TNG300-1-Dark simulation (A. Pillepich et al. 2018).

The paper is structured as follows: In Section 2, we present the photometric catalogs utilized in this study and the sample selection. Section 3 describes the methodology used to measure the galaxy clustering and abundance (SMF) ultimately used to constrain the HOD models. The adopted HOD modeling along with our new approach of fitting individual *z*-bins connected through smooth continuity priors is detailed in Section 4. This is followed by our SHMR results presented in Section 5. In Section 6, we discuss the intuition gained based on the results about the assembly of centrals and satellites and their host halos. We conclude with a summary in Section 7.

Throughout this paper, we adopt a flat $\Lambda$CDM cosmology with the following parameter values at $z = 0$ (denoted by the subscript "0"): Hubble constant, $H_0 = 70$ km s$^{-1}$ Mpc$^{-1}$; matter density, $\Omega_{m,0} = 0.3$ with a baryon fraction of 15.8% (Planck Collaboration et al. 2020) giving baryon density, $\Omega_{b,0} = 0.0474$; dark energy density, $\Omega_{\Lambda,0} = 0.7$; CMB temperature, $T_{\rm cmb,0} = 2.7255$ K. The masses are always quoted with the implicit dependence on the dimensionless scaled Hubble constant, $h = 0.7$ (complying with the $H_0$ given above): $M_\odot h^{-1}$ for dark matter halo masses and $M_\odot h^{-2}$ for the stellar masses. Furthermore, in the derivation of stellar population properties, we assumed the Chabrier initial mass function (IMF; G. Chabrier 2003) and the Calzetti dust attenuation law (D. Calzetti et al. 2000).

## 2. Data

### 2.1. COSMOS and UDS Catalogs

For this study, we utilize two homogeneously prepared catalogs in the COSMOS (UVISTA (A. Muzzin et al. 2013) DR3, ultradeep stripes) and UDS (K. Zaidi et al. 2024) fields. These catalogs are based on deep, multiwavelength Optical-NIR (IRAC) photometry in 50 and 24 filters for COSMOS and UDS, respectively.

The photometric redshifts are derived using the spectral energy distribution (SED) template-fitting code `EAZY` (G. B. Brammer et al. 2008) for COSMOS and its python implementation `eazy-py` for UDS. The photo-*z* agrees well with the available spectroscopic redshifts. Within the redshift span probed in this study, $0.2 < z < 4.5$, for COSMOS, the normalized median absolute deviation, $\sigma_{\rm NMAD} = 0.00505$, and the catastrophic outlier percentage, $(\frac{\Delta z}{1 + z_{\rm spec}} > 0.15)$, is $\sim$1.60%. For the same redshift span, for UDS, $\sigma_{\rm NMAD} = 0.01242$, while the catastrophic outlier percentage is ∼2.38%. The details on the spectroscopic redshifts considered for these analyses are mentioned in K. Zaidi et al. (2024) for UDS and A. Muzzin et al. (2013) for COSMOS. As spectroscopic redshifts are dominated by bright objects, we also addressed this bias in K. Zaidi et al. (2024) for UDS by estimating the photo-z errors for the full catalog using the close-pair method (R. F. Quadri & R. J. Williams 2010). This method leverages the fact that galaxies that are close together in the sky have a high likelihood of being at the same redshift, as galaxies tend to cluster together. Unavoidably, some contributions to close pairs also come from random projections from different redshifts; however, since they are random, their contribution can be easily estimated and subtracted out to leave physically associated close pairs. The dispersion in the photo-z distributions of these true close pairs then allows estimating the error on the photo-$z$. Using this method on the UDS catalog in K. Zaidi et al. (2024), we found the photo-$z$ errors to be within ∼0.7%−4.5% depending on redshift, stellar mass, and $K$-band magnitude considered.

The stellar masses along with other stellar population parameters were derived using the Dense Basis code (K. Iyer & E. Gawiser 2017; K. G. Iyer et al. 2019) which uses nonparametric star formation histories (SFHs) that are able to robustly capture different types of stellar populations giving unbiased estimates of stellar population properties as opposed to the codes implementing a singular functional form for SFH (L. Mowla et al. 2022). As shown in K. Zaidi et al. (2024), the `FAST` code (M. Kriek et al. 2009), implementing a singular functional form for the SFH, a delayed-tau exponential, provides systematically lower stellar masses by a median offset of ∼0.4 dex at $M_* \lesssim 10^{10} M_\odot$. Therefore, we also report the effect of this stellar mass offset in Section 5.4. In both cases, we assumed the G. Chabrier (2003) IMF and the D. Calzetti et al. (2000) dust attenuation law.

### 2.2. Sample Selection

To measure the clustering and the SMF reliably, we select galaxies with robust photometry meaning that they are (i) identified as a galaxy (not a star), (ii) do not fall on contaminated pixels which can be due to the proximity to nearby bright stars or other image artifacts, and are brighter than the 90% completeness limit in the detection band (see below for details). This selection was achieved using the masks generated for bright stars/artifacts/bad pixels and the following catalog cuts. For COSMOS, `use`= 1, `star`= 0, `contamination`= 0, `nan_contam` < 3, `K_flag` ⩽ 3, and 15 < total mag$_{K_s}$ [AB] < 24.5 (90% completeness), and similarly for UDS, `use_phot` = 1, `flag_Kuds` ⩽ 3, and total mag$_K$ [AB] < 24.3 (90% completeness). Additionally for UDS, we also avoid noisy data found along the four sides of the image. After such exclusions, we are left with the effective areas of 2954.51 arcmin$^2$ (∼0.82 deg$^2$) in COSMOS and 2828.25 arcmin$^2$ (∼0.79 deg$^2$) in UDS, as shown in Figure 1, which equals a combined effective area of ∼1.61 deg$^2$.

We determined the completeness for both COSMOS and UDS by calculating the recovered detections of injected stars in the images as a function of magnitudes in the detection bands, $K_s$ and $K$, for COSMOS and UDS, respectively. Since we estimated the completeness by injecting stars which are point sources, instead of galaxies which are extended sources, it is valid to ponder if the true completeness should be worse than the one estimated here. However, as we are dealing with ground-based high-redshift catalog in which galaxies are not much more resolved than point sources with few low-surface-brightness galaxies, such a simplification should not lead to highly biased estimates of the completeness of galaxies. We injected the stars at random locations on the images allowing for overlap with other sources to mimic the original images where some overlap is natural. The 90% completeness in the detection $K_s$-band for COSMOS is 24.5 AB, whereas the 90% completeness in the detection $K$-band for UDS is 24.3 AB.

Then, we used these estimates of 90% $K_s/K$ completeness limits to derive the stellar mass completeness for both fields separately following the method outlined in S. Juneau et al. (2005) and L. Pozzetti et al. (2010). We scaled the stellar masses of galaxies as a function of redshift to the masses they would have if they were at the limiting magnitudes mentioned above. This gives a representative distribution of mass-to-light (M/L) ratios for our two fields at the limiting magnitudes. We chose the 80th percentile of this distribution as a function of redshift as the stellar mass limit (80% stellar mass completeness) above which most galaxies are observable.

Finally, we restrict our combined sample (UDS+COSMOS) in this study to the higher of the two 80% stellar mass completeness limits in UDS and COSMOS which lie very close to each other as shown in Figure 2. Our final sample, divided into nine $z$-bins within $0.2 < z < 4.5$ and further into stellar mass threshold ($M_*^t$) samples, is marked with black lines ("ladders") in Figure 2 and listed in Table 1. Note that the subsamples are cumulative as opposed to binned in stellar mass, i.e., each stellar mass threshold sample contains all the galaxies within the stellar mass threshold samples below it.

The complete stellar mass indeed evolves within each $z$-bin, as is visible in the nonhorizontal completeness curve. Therefore, to fix our lowest stellar mass threshold within each $z$-bin, we chose it to be at the intersection of the completeness curve with the highest redshift end of any $z$-bin. These are the bottom-right corners of the black "ladders" within each $z$-bin in Figure 2. By doing so, we ensure that any $z$-bin sample is at least 80% complete throughout the redshift thickness of its bin. Within each $z$-bin, the samples become more and more complete as the redshift decreases.

Now, we address the consequence of our choice of the lower limit in stellar mass completeness—80% instead of 95%. By choosing an 80% completeness criteria, we miss some galaxies with the highest $M_*/L_*$ ratios (old and/or dusty galaxies) lying close to the bottom-right corners of the $z$-bin samples (see Figure 2). We determined that the 95% completeness curve is higher in stellar mass than the 80% completeness curve with a median increase of ∼0.19 dex for UDS and ∼0.14 dex for COSMOS. This incompleteness typically only affects the bottom-right corner of the lowest stellar mass threshold samples, depending on redshift, and since we jointly model all the stellar mass threshold samples using the L11 HOD model, we expect the consequence of this choice on the final results to be subservient to the errors on photo-$z$ and stellar mass. Additionally, it should be noted that as the dynamical range in stellar mass of our sample shrinks with redshift due to worsening completeness, our HOD constraints are increasingly informed via only massive galaxies going from low to high redshift. We highlight the stellar masses directly probed by our analysis in the figures showing results where appropriate.

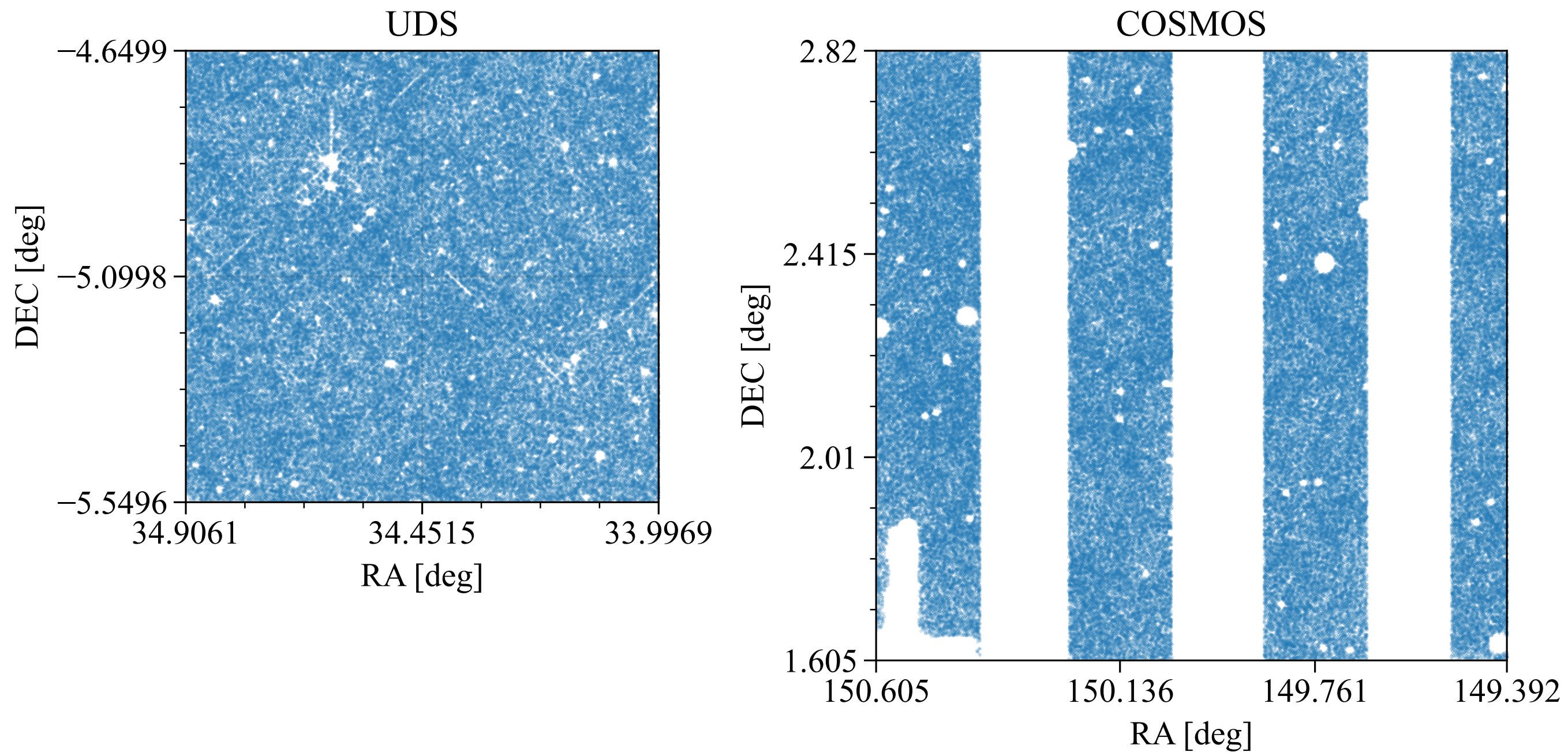

**Figure 1.** The final sample of galaxies selected in UDS ($N = 143{,}549$) and COSMOS ($N = 166{,}826$) is shown in blue points. The irregularly shaped white patches within both fields represent masked areas due to bright stars and other artifacts. In COSMOS, only ultradeep stripes of the UltraVISTA DR3 survey are utilized, while the shallower deep stripes, visible in white, are excluded. The relative dimensions of the panels highlight the relative R.A./Decl. span of the two fields. The effective areas of UDS and COSMOS are $\sim$0.79 deg$^2$ and $\sim$0.82 deg$^2$, respectively. This amounts to a total effective area of $\sim$1.61 deg$^2$.

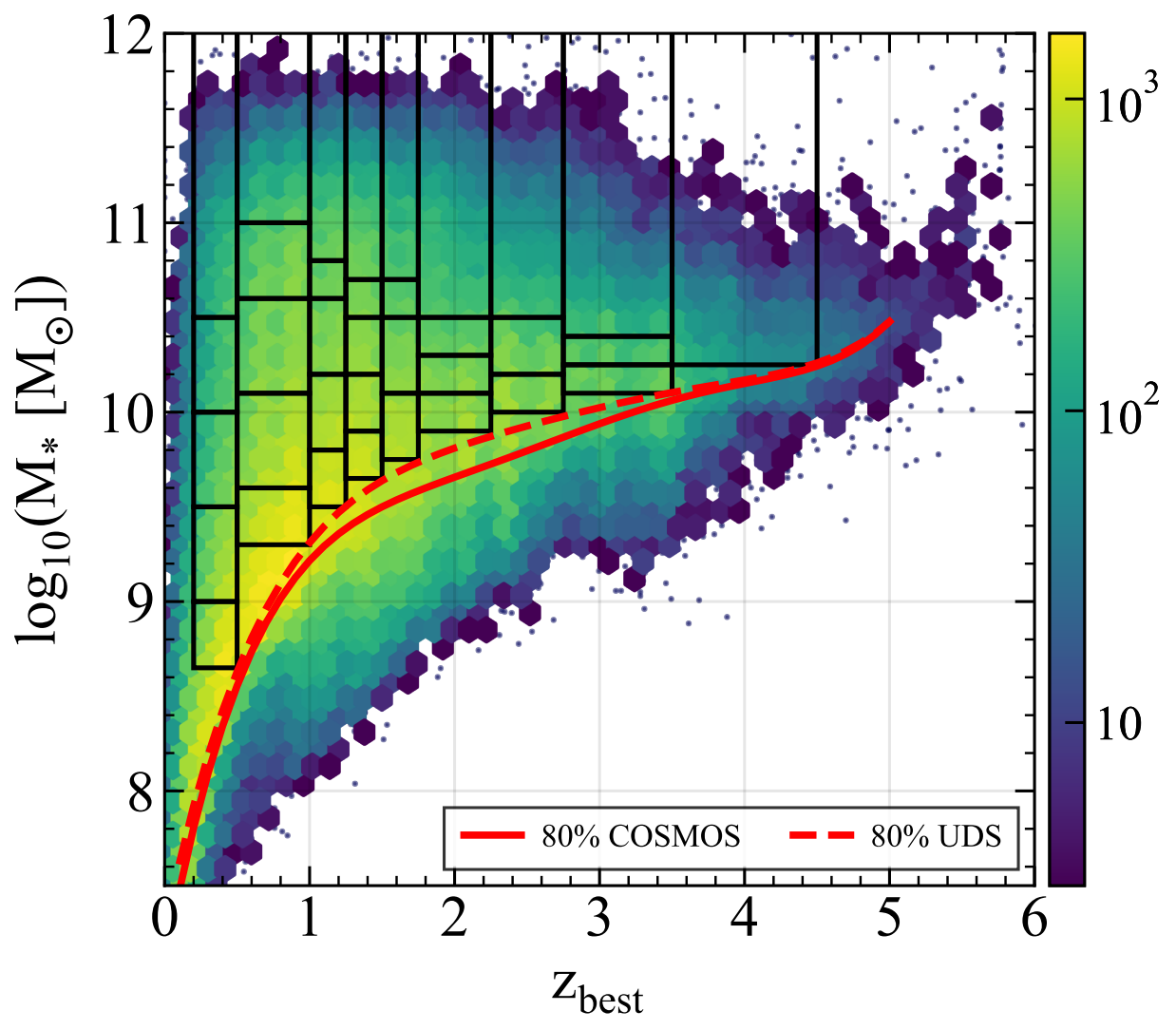

**Figure 2.** Distribution of the stellar mass versus redshift of the galaxies in COSMOS and UDS along with the selected subsamples (black ladders) for this study. The rungs of the ladders show the mass thresholds (for clustering) and mass bins (for SMF) of the subsamples. The right edge of the bottom rungs of the ladders at each redshift corresponds to the higher of the solid red and dashed red curves, representing the 80% stellar mass completeness of COSMOS and UDS, respectively.

## 3. Measurements

### 3.1. Two-point Angular Correlation Function

We compute the two-point angular correlation function, $w_{\rm obs}(\theta)$, for the stellar mass threshold samples which is essentially the excess probability above Poisson of finding a galaxy–galaxy pair as a function of angular scale, $\theta$. To do so, we use the popular S. D. Landy & A. S. Szalay (1993) estimator shown in Equation (1) within the python-based pair counter engine, CORRFUNC [12] (M. Sinha & L. H. Garrison 2020) implemented within the python utility wrapper, pycorr. [13] In Equation (1), DD, DR, and RR indicate the number of data–data, data–random, and random–random pairs. All of these pairs are normalized by the total number of possible pairs, as shown in Equation (1), where $n_{\rm D}$ and $n_{\rm R}$ indicate the number of points in the data and random catalogs, respectively.

$$w_{\rm obs}(\theta) = \frac{\left[\frac{\mathrm{DD}(\theta)}{n_{\rm D}^2} - 2\frac{\mathrm{DR}(\theta)}{n_{\rm D}n_{\rm R}} + \frac{\mathrm{RR}(\theta)}{n_{\rm R}^2}\right]}{\left[\frac{\mathrm{RR}(\theta)}{n_{\rm R}^2}\right]}. \quad (1)$$

The random catalog is created using exactly the same survey geometry mapped out by the galaxies considered in this study avoiding areas around bright stars, etc. Moreover, the number of points in the random catalogs is at least $\sim$65 times more than the galaxies in any stellar mass threshold sample ensuring that $w_{\rm obs}(\theta)$ is properly measured. We measure the $w_{\rm obs}(\theta)$ in 13 angular separation bins within $-3 < \log_{10}(\theta[\mathrm{deg}]) < -0.6$, which corresponds to a maximum separation of just over 0.25 deg. As a result we probe a maximum transverse comoving scale of $\sim$6 Mpc in our lowest $z$-bin, which increases to $\sim$32 Mpc for our highest $z$-bin.

We estimate the errors on $w_{\rm obs}(\theta)$ by calculating the square root of the diagonal elements of the covariance matrix capturing the covariance among all angular scales. To estimate the covariance matrix, we use the "delete-one" jackknife method (J. V. Wall & C. R. Jenkins 2012) wherein we subdivide the whole region of our sample footprint into 50 equal-sized patches. Then for each mass threshold sample, we compute the correlation functions for 50 realizations each time with a unique patch removed. Finally, for each subsample, the covariance matrix, $C_{ij}$, is computed as follows:

$$C_{ij} = \frac{N_r - 1}{N_r} \sum_{r=1}^{N_r} (w_r(\theta_i) - \tilde{w})^T (w_r(\theta_j) - \tilde{w}), \quad (2)$$

where $N_r$ is the number of realizations, $\tilde{w}$ is the average correlation function of all realizations, and $i$ and $j$ indices label the angular separation bins.

[12] https://github.com/manodeep/Corrfunc
[13] https://github.com/cosmodesi/pycorr

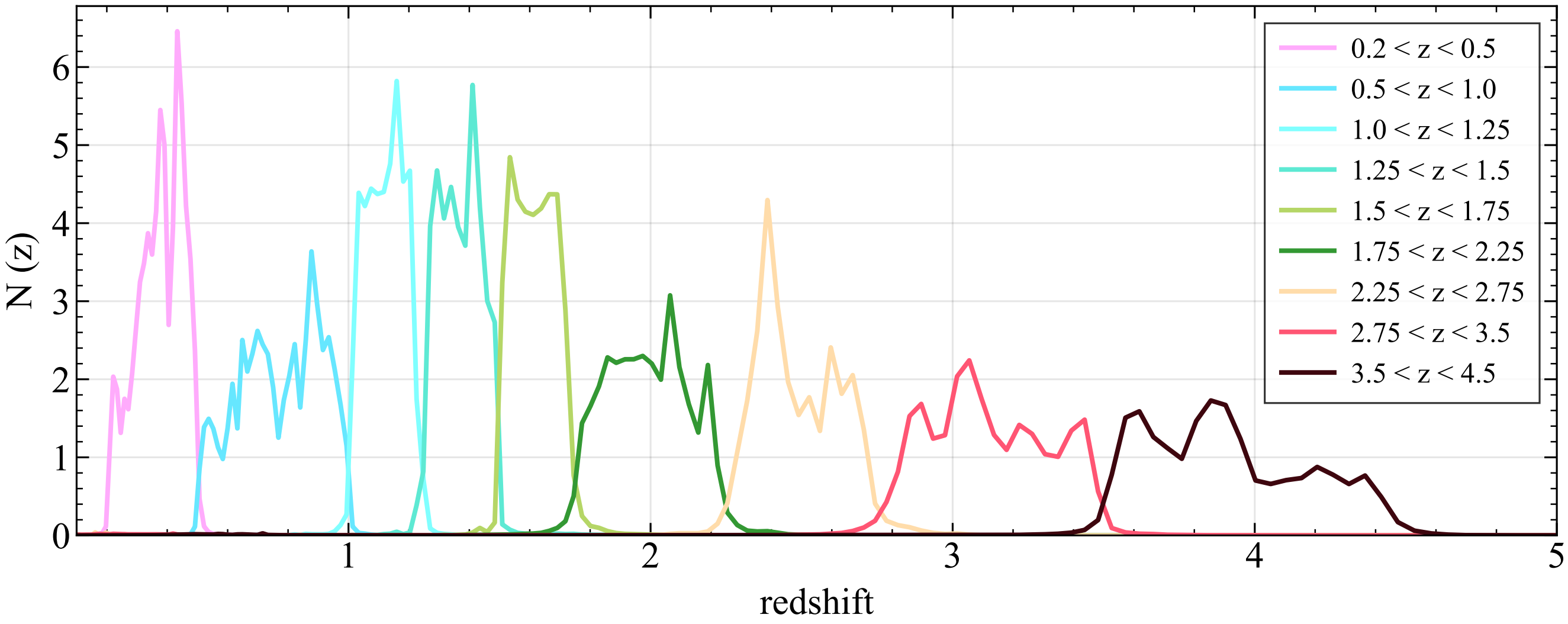

**Figure 3.** Redshift distribution, $N(z)$, for each of the nine $z$-bins used in this study, generated by stacking the galaxy redshift probability distributions given by `eazy-py` above the lowest stellar mass threshold within each $z$-bin.

**Table 1**
Subsamples Used in This Study

| $z$-bin | $\log_{10}(M_*^t[M_\odot])$ | $N_{\rm COSMOS}$ | $N_{\rm UDS}$ | $N_{\rm total}$ |
|---|---|---|---|---|
| $0.2 < z < 0.5$ | 8.65 | 12600 | 8912 | 21,512 |
| | 9.0 | 8927 | 6077 | 15,004 |
| | 9.5 | 5343 | 3645 | 8988 |
| | 10.0 | 3175 | 2166 | 5341 |
| | 10.5 | 1697 | 1091 | 2788 |
| $0.5 < z < 1.0$ | 9.3 | 26225 | 20926 | 47,151 |
| | 9.6 | 19652 | 15027 | 34,679 |
| | 10.1 | 11480 | 8470 | 19,950 |
| | 10.6 | 5699 | 4149 | 9848 |
| | 11.0 | 2060 | 1373 | 3433 |
| $1.0 < z < 1.25$ | 9.5 | 8628 | 8177 | 16,805 |
| | 9.8 | 6335 | 5826 | 12,161 |
| | 10.2 | 3893 | 3567 | 7460 |
| | 10.6 | 2257 | 2001 | 4258 |
| | 10.8 | 1572 | 1318 | 2890 |
| $1.25 < z < 1.5$ | 9.65 | 7549 | 9151 | 16,700 |
| | 9.9 | 5656 | 6900 | 12,556 |
| | 10.2 | 3805 | 4833 | 8638 |
| | 10.5 | 2403 | 3093 | 5496 |
| | 10.7 | 1669 | 2152 | 3821 |
| $1.5 < z < 1.75$ | 9.75 | 4268 | 5365 | 9633 |
| | 10.1 | 2677 | 3470 | 6147 |
| | 10.5 | 1317 | 1945 | 3262 |
| | 10.7 | 864 | 1417 | 2281 |
| $1.75 < z < 2.25$ | 9.9 | 6925 | 6196 | 13,121 |
| | 10.1 | 5280 | 4568 | 9848 |
| | 10.3 | 3713 | 3243 | 6956 |
| | 10.5 | 2485 | 2312 | 4797 |
| $2.25 < z < 2.75$ | 10.0 | 4820 | 4004 | 8824 |
| | 10.2 | 3110 | 2894 | 6004 |
| | 10.5 | 1298 | 1491 | 2789 |
| $2.75 < z < 3.5$ | 10.1 | 3772 | 3736 | 7508 |
| | 10.25 | 2599 | 2549 | 5148 |
| | 10.4 | 1533 | 1608 | 3141 |
| $3.5 < z < 4.5$ | 10.25 | 931 | 1256 | 2187 |

As we measure $w_{\rm obs}(\theta)$ from limited-sized patches on the sky (total effective area $\sim$1.61 deg$^2$), it is bound to be underestimated by a constant integral constraint (IC) factor (E. J. Groth & P. J. E. Peebles 1977). The IC correction scales inversely with the field size and directly with the clustering strength as it is equal to the fractional variance on galaxy counts (D. A. Wake et al. 2011). We describe how we deal with this correction in Section 4.6.

### 3.2. Redshift Distributions

In order to convert the modeled two-point correlation function in three dimensions $\xi(r)$ into angular correlation functions, $w_{\rm model}(\theta)$ (as discussed in Section 4.6), we need the redshift distributions, $N(z)$, for each of the $z$-bins in our sample. We derived them by stacking the `eazy-py` output redshift probability distributions, the $p(z)$ of galaxies above the lowest stellar mass threshold within each $z$-bin, ensuring that the 16th/84th percentiles of the $p(z)$ lie within the limits of the considered $z$-bin. The final $N(z)$ is shown in Figure 3, where the large-scale structure can be easily gleaned through its jagged nature with sharp spikes and dips.

### 3.3. Stellar Mass Function

We derive the SMFs in the UDS and COSMOS fields above the lowest stellar masses threshold at each $z$-bin as described in Section 2.2, defined in Table 1, and shown in Figure 2. As we only compute SMFs for essentially complete samples, we do not need to implement the usual $1/V_{\rm max}$ technique (M. Schmidt 1968) or similar, which is useful when not all of the galaxies within a stellar mass bin are observable within the full volume spanned by a given $z$-bin and survey area.

The Poisson errors on the SMFs were estimated using N. Gehrels (1986), which is especially useful for small number counts at the high-mass end. We also calculated cosmic variance errors for each of the two fields separately following B. P. Moster et al. (2011) and then combined them using the volume-weighted average technique described in the same paper. Briefly, to calculate the cosmic variance in the individual fields, we used their Equation (5): $\sigma_v^2\ (M_*, z) = b^2\ (M_*, z)\ \sigma_{dm}^2\ (z)$, where $b$ is the galaxy bias that depends on stellar mass and redshift mentioned therein, whereas $\sigma_{dm}^2$ is the cosmic variance error on dark matter based on redshift and the dimensions of the survey area calculated using the public QUICKCV code (J. A. Newman & B. P. Moster 2014). We derived the cosmic variance for the combined sample from the two fields using the individual field's cosmic variance and survey area as in Equation 9 of B. P. Moster et al. (2011). Finally, the total uncertainties on the SMFs were calculated by adding the Poisson errors ($\sigma_{\rm Poisson}$), errors due to cosmic variance ($\sigma_{\rm CV}$), and a fixed 3% error attributed to the uncertainty in the survey area ($\sigma_{\rm Area}$) in quadrature. We show the individual and combined SMFs of the two fields in Appendix A. Additionally, we also tabulate the combined SMFs and the associated percentage errors due to all three sources mentioned above in Appendix A.

## 4. Modeling the HOD

We derive the SHMR under the framework of HOD which models the average number of galaxies selected above a stellar mass threshold as a function of halo mass. There are many implementations of it in the literature (e.g., Z. Zheng et al. 2005, 2007; I. Zehavi et al. 2005, 2011; A. Leauthaud et al. 2011; Y. Zu & R. Mandelbaum 2015), most of which at least split the contribution from the central galaxy and satellite galaxies to the average total occupation number at a given halo mass. The central galaxy is the most massive galaxy dominating the potential well of a halo and, depending on the halo mass, it might be surrounded by one or more lower-mass satellite galaxies.

The clustering measurements are an amalgamation of galaxy–galaxy (satellite–satellite or central–satellite) pairs within the same halos and therefore dominant at smaller angular scales, and the galaxy–galaxy (central–central or central–satellite) pairs between different halos that contribute more at larger scales. These are the so-called "one-halo" (pairs from the same halos) and "two-halo" (pairs from different halos) contributions to the total clustering which result in a break in total clustering at a certain angular scale where the clustering strength of "one-halo" drops and "two-halo" picks up (I. Zehavi et al. 2004).

### 4.1. SHMR of Central Galaxies

In this study, we utilize the HOD prescribed by L11 which starts by assuming a functional form for SHMR, $f_{\rm SHMR}(M_{\rm h})$, which gives the mean-log stellar mass, $\langle \log_{10}(M_*(M_{\rm h}))\rangle$ for central galaxies as a function of a fixed halo mass, $M_h$. L11 adapted the $f_{\rm SHMR}(M_{\rm h})$ from P. S. Behroozi et al. (2010). For convenience, $f_{\rm SHMR}(M_{\rm h})$ can be inversely defined as follows:

$$\log_{10}\left(f_{\rm SHMR}^{-1}(M_*)\right) = \log_{10}(M_h)$$
$$= \log_{10}(M_1) + \beta \log_{10}\left(\frac{M_*}{M_{*,0}}\right) + \frac{\left(\frac{M_*}{M_{*,0}}\right)^{\delta}}{1 + \left(\frac{M_*}{M_{*,0}}\right)^{-\gamma}} - \frac{1}{2}, \quad (3)$$

where $M_1$ governs the characteristic halo mass, where the higher its value, the more massive the host halo for a given stellar mass; $M_{*,0}$, in turn, governs the characteristic stellar mass, such that increasing it results in less massive host halos assigned to a given stellar mass; $\beta$ controls the power-law slope at the low-mass end; $\delta$ controls the subexponential slope on the high-mass end; finally, $\gamma$ dictates how sharp the transition is between the low-mass and high-mass regimes. The effect of varying each of these parameters on the shape of SHMR can also be seen in Figure 1 in L11.

### 4.2. Scatter in Stellar Mass at a fixed Halo Mass

At a fixed halo mass there is still expected to be a scatter in the stellar mass, which arises from the intrinsic scatter in the Universe and the scatter introduced by measurement errors due to SED modeling, photo-z uncertainties, and the errors in the observed magnitudes (P. S. Behroozi et al. 2010; L11). L11 captures this stochasticity by assuming a log-normal distribution describing the probability distribution function of $\log_{10}(M_*)$ at a fixed halo mass. The standard deviation of this distribution, $\sigma_{\log M_*}$, i.e., the log-normal scatter in the stellar mass at a fixed halo mass is then fitted as a free parameter.

The data at the high-mass end mainly tend to constrain the $\sigma_{\log M_*}$ (A. Leauthaud et al. 2012). It is because first, the low-mass data leave a weaker imprint on the data (A. Leauthaud et al. 2012), and second, the low-mass data below the completeness limit are excluded in our analysis as opposed to, e.g., Y. Zu & R. Mandelbaum (2015), where a modified modeling technique was employed to incorporate the mass regime where the data are incomplete. Therefore, the inclusion of halo mass dependence in $\sigma_{\log M_*}$ in our case does not matter in any meaningful way. In fact, A. Leauthaud et al. (2012) considered two models, one in which $\sigma_{\log M_*}$ is a constant and another in which it varies as a function of stellar mass. They found no meaningful differences in the resulting HOD parameter fit and the SHMR model. As a consequence, we have assumed $\sigma_{\log M_*}$ to be independent of halo mass in this work, i.e., $\sigma_{\log M_*}$ is considered to be constant for all halo masses within a $z$-bin.

### 4.3. Mean Occupation of Central Galaxies

To compute the mean occupation of central galaxies, we can integrate the conditional SMF for centrals, $\Phi_{\rm cen}(M_*|M_h)$, which is the number of centrals with stellar mass within $M_* \pm dM_*/2$ given a fixed halo mass. For instance, for a sample of centrals with a stellar mass threshold, $M_*^t$, the mean occupation can be calculated as follows:

$$\langle N_{\rm cen}(M_h|M_*^t)\rangle = \int_{M_*^t}^{\infty} \Phi_{\rm cen}(M_*|M_h)\, dM_* \quad (4)$$

In L11, $\Phi_{\rm cen}(M_*|M_h)$ is defined as log-normal distribution (see their Equation (1)) which means that $\langle N_{\rm cen}(M_h|M_*^t)\rangle$ can be

derived analytically. The final expression for $\langle N_{\rm cen}(M_h|M_*^t)\rangle$ using $f_{\rm SHMR}(M_h)$ and a constant $\sigma_{\log M_*}$ then becomes (see L11 for the shape of $\langle N_{\rm cen}(M_h|M_*^t)\rangle$ at different $M_*^t$):

$$\langle N_{\rm cen}(M_h|M_*^t)\rangle = \frac{1}{2}\left[1 - {\rm erf}\left(\frac{\log_{10}(M_*^t) - \log_{10}(f_{\rm SHMR}(M_h))}{\sqrt{2}\,\sigma_{\log M_*}}\right)\right], \quad (5)$$

where the error function, erf is defined by

$${\rm erf}(x) = \frac{2}{\sqrt{\pi}}\int_0^x e^{-t^2}dt.$$

The alternate commonly employed expression for the mean occupation function of centrals in the literature, e.g., as in Z. Zheng et al. (2005), can be derived from Equation (5) if (i) $f_{\rm SHMR}(M_h)$ is assumed to be a power law at all masses, and (ii) a minimum halo mass, $M_{\rm min}$, which corresponds to $M_*^t$, is introduced, i.e., $M_{\rm min} \equiv f_{\rm SHMR}^{-1}(M_*^t)$ (L11). Even though this common version agrees well with Equation (5) at low masses ($M_{\rm min} \lesssim 10^{12}\ M_\odot$), the difference between the two grows between 10% and 40% at higher masses (L11) as the L11 SHMR deviates from the power law and becomes subexponential at high masses. As observations confirm that SHMR rises faster than a power law at high masses, it is crucial to use an expression for $\langle N_{\rm cen}\rangle$ that explicitly incorporates a shift from power law to subexponential at higher masses.

### 4.4. Mean Occupation of Satellite Galaxies

The model for the mean occupation of satellites is expected to follow a power law at high masses with an exponential sharp cut-off at low masses (A. A. Berlind et al. 2003; A. V. Kravtsov et al. 2004; Z. Zheng et al. 2005). Therefore, the expression for $\langle N_{\rm sat}\rangle$ can be written as in Equation (6) where the $\alpha_{\rm sat}$ is the power-law slope with $M_{\rm sat}$ as the power-law amplitude. $M_{\rm cut}$ defines the scale for the exponential cut-off. Moreover, halos are not expected to have satellite galaxies before having a central galaxy. This is ensured by scaling to the satellite occupation to $\langle N_{\rm cen}\rangle$.

$$\langle N_{\rm sat}(M_h|M_*^t)\rangle = \langle N_{\rm cen}(M_h|M_*^t)\rangle\left(\frac{M_h}{M_{\rm sat}}\right)^{\alpha_{sat}}\exp\left(\frac{-M_{\rm cut}}{M_h}\right). \quad (6)$$

As it turns out according to the observational studies, $M_{sat}$ is proportional to $M_{\rm min}$, where $M_{\rm min}$ is the inverse of the SHMR, i.e., $M_{\rm min}{\equiv}f_{\rm SHMR}^{-1}(M_*^t)$, where $f_{\rm SHMR}^{-1}(M_*^t)$ is assumed to be a power law (L11). For added flexibility, instead of $M_{sat}$ and $M_{cut}$ simply scaling with $f_{\rm SHMR}^{-1}(M_*^t)$, L11 defines $M_{\rm sat}$ and $M_{\rm cut}$ as power-law functions of $f_{\rm SHMR}^{-1}(M_*^t)$ as shown in Equations (7) and (8), respectively.

$$\frac{M_{\rm sat}}{10^{12}M_\odot} = B_{\rm sat}\left(\frac{f_{\rm SHMR}^{-1}(M_*^t)}{10^{12}M_\odot}\right)^{\beta_{\rm sat}}, \quad (7)$$

$$\frac{M_{\rm cut}}{10^{12}M_\odot} = B_{\rm cut}\left(\frac{f_{\rm SHMR}^{-1}(M_*^t)}{10^{12}M_\odot}\right)^{\beta_{\rm cut}}. \quad (8)$$

Furthermore, we can easily calculate the fraction of stellar mass locked up in satellites, $f_{\rm sat}$, by dividing the mean number density of satellites, $\tilde{n}_{g,\ \rm sat}$ with the mean total (central + satellites) number density, $\tilde{n}_g$, as shown in Equation (9) below. Here, the mean number densities are calculated by integrating the HMF, $dn/dM_h$, along with the desired mean occupation function.

$$f_{sat} = \frac{\tilde{n}_{g,\ sat}}{\tilde{n}_g} = \frac{\int dM_h \frac{dn}{dM_h}\langle N_{\rm sat}(M_h|M_*^t)\rangle}{\int dM_h \frac{dn}{dM_h}\langle N_{\rm total}(M_h|M_*^t)\rangle} \quad (9)$$

### 4.5. Total Stellar Mass in Halos

The L11 HOD model also allows us to conveniently calculate the total stellar mass in galaxies as a function of halo mass and the contributions from central and satellite galaxies. The total stellar mass within a stellar mass bin can be calculated by integrating the conditional SMFs of centrals and satellites multiplied by $M_*$ with the stellar mass range as the limits of integration as shown in the second line of Equation (10). Then, using the fact that integrating the conditional SMF of centrals/satellites gives the occupation number of centrals/satellites and the integration by parts rule, we can further simplify the integration as shown in the third and fourth lines of Equation (10):

$$\begin{aligned} M_*^{\rm tot}&\left(M_h|M_*^{t_1}, M_*^{t_2}\right) \\ &= \int_{M_*^{t_1}}^{M_*^{t_2}}[\Phi_c(M_*|M_h) + \Phi_s(M_*|M_h)]M_* dM_* \\ &= \int_{M_*^{t_1}}^{M_*^{t_2}}\langle N_{\rm cen}(M_h|M_*)\rangle dM_* - [\langle N_{\rm cen}(M_h|M_*)\rangle M_*]_{M_*^{t_1}}^{M_*^{t_2}} \\ &\quad + \int_{M_*^{t_1}}^{M_*^{t_2}}\langle N_{\rm sat}(M_h|M_*)\rangle dM_* - [\langle N_{\rm sat}(M_h|M_*)\rangle M_*]_{M_*^{t_1}}^{M_*^{t_2}}, \end{aligned} \quad (10)$$

where the first term (third line) reflects the contribution from the central galaxies and the second term (fourth line) corresponds to the satellite galaxies.

### 4.6. Two-point Angular Correlation Function from the HOD

To derive the modeled two-point angular correlation function, $w_{\rm halomod}(\theta)$, within the halomod code (S. G. Murray et al. 2021), we start from the model galaxy real-space correlation function, $\xi_{gg}(r)$, the derivation of which is described in detail in S. G. Murray et al. (2021). Briefly, $\xi_{gg}(r)$ is connected to the galaxy power spectrum, $P_{gg}(k)$, via order-1/2 Hankel transform:

$$\xi_{gg}(r) = \frac{1}{2\pi^2}\int_0^\infty P_{gg}(k)k^2 j_0(kr)dk, \quad (11)$$

with $j_0$ being the zeroth-order spherical Bessel function, $j_0(x) = \frac{\sin x}{x}$. $P_{gg}(k)$ is further divided into contributions from the one-halo (1h) and two-halo (2h) pairs: $P_{gg}(k) = P_{gg}^{1h}(k) + P_{gg}^{2h}(k)$, and correspondingly, $\xi_{gg}(r) = [1 + \xi_{gg}^{1h}(r)] + \xi_{gg}^{2h}(r)$. $P_{gg}^{1h}(k)$ and $P_{gg}^{2h}(k)$ are defined as follows:

$$P_{gg}^{1h}(k) = \int dM_h \frac{dn}{dM_h}u(k|M_h)\left[\langle N_{\rm sat}\rangle^2 u(k|M_h) + 2\langle N_{\rm sat}N_{\rm cen}\rangle\right], \quad (12)$$

$$P_{gg}^{2h}(k) = P_m(k)\left[\int dM_h \frac{dn}{dM_h} u(k|M_h) \frac{\langle N_{\text{total}}\rangle}{\tilde{n}_g} b(M_h)\right]^2. \quad (13)$$

In Equations (12) and (13), $u(k|M_h)$ is the halo profile's mass-normalized Fourier transform; $P_m(k)$ is the matter power spectrum; $\tilde{n}_g$ is the mean number density of galaxies given calculated as in Equation (9); and $b(M_h)$ is the scale-independent halo bias from J. L. Tinker et al. (2010).

From $\xi(r)$, we derive the angular correlation function within the `halomod` code (S. G. Murray et al. 2021) using Limber's equation (D. N. Limber 1954) as presented in C. Blake et al. (2008):

$$w_{\text{halomod}}(\theta) = 2\int_0^\infty dx f^2(x) \int_0^\infty du \xi(r = \sqrt{u^2 + x^2\theta^2}) \quad (14)$$

$$f(x) = \frac{p(z)}{dx/dz(z)}, \quad (15)$$

where $p(z)$ is the redshift distribution which we deduce from the stacking of the individual galaxy's redshift probability density functions within a $z$-bin. $x$ is the comoving radial distance from the median redshift, and $f(x)$ defines the overall radial distribution.

As mentioned in Section 3.1, the measured angular correlation function, $w_{\text{obs}}(\theta)$, is underestimated by a constant additive IC factor. Instead of adding this correction to $w_{\text{obs}}(\theta)$, we subtract it from $w_{\text{halomod}}(\theta)$ to get the model angular correlation function, $w_{\text{model}}(\theta)$, i.e., $w_{\text{model}}(\theta) = w_{\text{halomod}}(\theta)$ - IC. We estimate IC as in L. Infante (1994) and N. Roche & S. A. Eales (1999):

$$\text{IC} = \frac{\sum_i w_{\text{halomod}}(\theta_i) RR(\theta_i)}{\sum_i RR(\theta_i)}, \quad (16)$$

where we run the $\theta_i$ out to much larger separations (∼4 deg) than while measuring/modeling the angular correlation function itself (only out to ∼0.25 deg) to ensure that we have counted all of the $RR$ pairs. While fitting we can now compare $w_{\text{obs}}(\theta)$ and $w_{\text{model}}(\theta)$ directly (see Section 4.9) as both of them are lower by the same IC factor.

### 4.7. Stellar Mass Functions from the HOD

The SMF corresponding to a chosen set of HOD parameter values can be calculated using the occupation function and the HMF, $dn/dM_h$. Now, Equations (5) and (6) define the occupation functions with respect to a stellar mass threshold; however, the SMF is a series of values of number density per unit volume in bins of stellar mass. This means that we need the occupation function defined in bins of stellar masses instead of in thresholds of stellar mass. Fortunately, the occupation function within a stellar mass bin is simply the difference between the two stellar mass threshold-defined occupation functions at the mass thresholds $M_*^{t_1}$ and $M_*^{t_2}$ defining the bin edges:

$$\langle N(M_h|M_*^{t_1}, M_*^{t_2})\rangle = \langle N(M_h|M_*^{t_2})\rangle - \langle N(M_h|M_*^{t_1})\rangle, \quad (17)$$

where $\langle N\rangle$ represents the mean occupation function of the centrals, satellites, or total depending on whichever is desired. For a stellar mass bin defined by $\Delta\log_{10}M_* = M_*^{t_2} - M_*^{t_1}$, the modeled SMF, $\Phi_{\text{model}}(M_*^{t_1}, M_*^{t_2})$, can then be calculated as:

$$\Phi_{\text{model}}(M_*^{t_1}, M_*^{t_2})\Delta log_{10}M_* = \int_0^\infty \left(\langle N(M_h|M_*^{t_1}, M_*^{t_2})\rangle \frac{dn}{dM_h}\right) dM_h, \quad (18)$$

where again $\langle N\rangle$ can be the mean occupation function of the centrals, satellites, or total depending on whichever SMF is desired.

### 4.8. Redshift Evolution of the HOD Parameters

As can be seen in Figure 2, the dynamical range in the stellar mass shrinks dramatically with redshift due to the widening range of incomplete stellar masses. The lowest stellar mass considered within the highest $z$-bin is $10^{10.3}$ $M_\odot$ about 2 orders of magnitude higher than considered at the lowest $z$-bin. As a result, it becomes challenging to constrain the low-mass slope (controlled by $\beta$) of the SHMR at high $z$. Due to the small dynamical range in stellar mass, the scatter in stellar mass at a fixed halo mass, $\sigma_{\log M_*}$, also proves harder to constrain.

Furthermore, even though the total effective survey area used in this study is large, it is not large enough to capture a high number of massive galaxies at all redshifts. Therefore, at all redshifts, it remains a struggle to constrain the $\delta$ and $\gamma$ parameters which control the slope of the SHMR at the massive end and the transition regime between low-mass and high-mass, respectively.

This motivated us to consider an alternate approach to fitting the HODs. All of the 11 HOD parameters at each $z$-bin can be better constrained if it is realized that the clustering and SMFs in the neighboring $z$-bins are not disconnected from each other but rather evolve from one to another. Consequently, the HOD parameters that produce the clustering and abundance should also evolve smoothly, rather than having sharp jumps between the neighboring $z$-bins. We implement this understanding by penalizing sharp changes in the HOD parameter values between the adjacent $z$-bins through the use of continuity ("smoothing") priors.

Consequently, instead of fitting the clustering and abundance data for each of the individual $z$-bins ("discrete-$z$" model hereafter), we jointly fit the data in all redshifts at once with the HOD evolving smoothly with redshift due to the continuity priors ("smooth-$z$" model hereafter). This approach allows us to leverage the wealth of information contained within the clustering and abundance measurements at lower redshift spanning wider dynamical ranges in stellar mass, by connecting them to higher redshift via smooth redshift evolution of the HOD parameters.

To implement this smoothing, we use Gaussian priors that peak when there is no change in a given HOD parameter between the adjacent $z$-bins. For the HOD parameters that are fitted in log-space, $M_{*,0}$, $M_1$, and $\sigma_{\log M_*}$, we impose the Gaussian continuity prior (centered at "0") on the difference in the HOD parameter values between the adjacent $z$-bins, e.g.,

**Table 2**
$\sigma$ of the Gaussian Priors on the Change in HOD Parameters between Adjacent Bins

| HOD Parameter | Range | $\sigma$ |
|---|---|---|
| $M_{*,0}$ | 10–12 | 0.065 |
| $M_1$ | 11.8–14 | 0.12 |
| $\beta$ | 0.1–0.9 | 0.06 |
| $\delta$ | 0.1–4 | 0.08 |
| $\gamma$ | 0–8 | 0.12 |
| $\sigma_{\log M_*}$ | 0.001–0.5 | 0.025 |
| $\alpha_{\rm sat}$ | 0.4–2 | 0.07 |
| $\beta_{\rm sat}$ | 0.01–4 | 0.09 |
| $B_{\rm sat}$ | 0–17 | 0.065 |
| $\beta_{\rm cut}$ | −1–8 | 0.145 |
| $B_{\rm cut}$ | 0–14 | 0.145 |

$\Delta \equiv M_1(z_x) - M_1(z_{x+1})$ as follows:

$$\text{prior}(\Delta) = \frac{e^{-\frac{1}{2}\left(\frac{\Delta}{\sigma}\right)^2}}{\sigma\sqrt{2\pi}} \tag{19}$$

where $x$ denotes any $x$th $z$-bin. Similarly, for all of the eight other HOD parameters that are not fitted in log space, we impose the continuity prior (centered at “1”) on the ratio of the HOD parameter values between the adjacent $z$-bins, e.g., $\Delta \equiv \frac{\beta(z_x)}{\beta(z_{x+1})}$ as follows:

$$\text{prior}(\Delta) = \frac{e^{-\frac{1}{2}\left(\frac{\Delta-1}{\sigma}\right)^2}}{\sigma\sqrt{2\pi}} \tag{20}$$

where $\sigma$ is the width of the Gaussian continuity prior decided such that the maximum range obtained with the “discrete-$z$” fits is allowed to occur over the whole redshift range. Their values of $\sigma$ along with the bounds on all the HOD parameters are tabulated in Table 2.

### 4.9. Fitting the HOD

Our primary results come from the “smooth-$z$” model in which we jointly fit the HOD parameters at the nine $z$-bins (described in 4.8) by minimizing the following $\chi^2$:

$$\chi^2 = \sum_z^Z \chi_w^2 + \chi_\Phi^2 \tag{21}$$

where

$$\chi_w^2 = \sum_i^{N_{M_*^t}} (\boldsymbol{w}_{\mathbf{obs}\,i} - \boldsymbol{w}_{\mathbf{model}\,i})^{\rm T}\, \boldsymbol{C}_{obs,\,i}^{-1}\, (\boldsymbol{w}_{\mathbf{obs}\,i} - \boldsymbol{w}_{\mathbf{model}\,i}) \tag{22}$$

and

$$\chi_\Phi^2 = \sum_j^{N_{M_*^b}} \left( \frac{\log_{10}(\Phi_{\rm obs,\ \it j}) - log_{10}(\Phi_{\rm model,\ \it j})}{log_{10}(\sigma_{\Phi_{\rm obs,\ \it j}})} \right)^2. \tag{23}$$

In Equation (22) above, the summing index $i$ runs over $N_{M_*^t}$ stellar mass threshold samples (shown in Table 1), and $C_{\rm obs,\ \it i}^{-1}$ is the inverse covariance matrix calculated using Equation (2). In Equation (23), the summing index $j$ runs over $N_{M_*^b}$ stellar mass bins used to evaluate the SMFs and $\sigma_{\Phi_{\rm obs,\ \it j}}$ is the total error on the observed SMF, $\Phi_{\rm obs}$, including contributions from both the Poisson statistics and the cosmic variance as described in Section 3.3. Finally, as we jointly fit all redshifts simultaneously, the index $z$ in the outer sums in both terms runs over all the $Z$ (= 9) $z$-bins.

In practice, we maximize the log-likelihood, $\ln(\mathcal{L}) = -\chi^2/2$, which is the same as minimizing the $\chi^2$ using the python package `emcee` (D. Foreman-Mackey et al. 2013), which implements the affine-invariant Markov Chain Monte Carlo (MCMC) ensemble sampler of J. Goodman & J. Weare (2010). As we jointly fit the HODs for all the $z$-bins, we have a total of 99 parameters = 11 HOD parameters at each $z$-bin × 9 $z$-bins to be fit. We run 250 walkers to fit them until the resulting chains are converged. We consider the chains converged if they pass the typical convergence criteria that their auto-correlation time, $\tau$ (number of iterations before the chain “forgets” where it started from), is at least 50 times smaller than the total number of iterations and that the change in chains is less than ∼1%. After rejecting a certain number of iterations as “burn-in,” we retrieved the marginalized posterior distributions from the chains. The medians of the posteriors are considered the best-fit values and the errors on them come from the 16th/84th percentiles.

To show the improvement in the HOD parameter fits due to the novel “smooth-$z$” model, we also perform the HOD fitting in the standard way (“discrete-$z$”) and compare the results in Section 5.2. For the “discrete-$z$” model, the $\chi^2$ formulation remains the same as in Equation (21) except for the outer sums running over all the $z$-bins which no longer apply. Furthermore, for the “discrete-$z$” model as it fits each $z$-bin separately, we now only have 11 HOD parameters per fit. We fit them using 60 walkers until the chains are converged according to the same criteria as mentioned above.

## 5. Results

### 5.1. Evolution of the Measured Clustering and Abundance

In Figure 4, we show our measurements (empty circles with error bars) of the two-point angular correlation function, $w_{\rm obs}(\theta)$ for all of the stellar mass threshold subsamples within each $z$-bin as defined in Table 1. The clustering of different stellar mass threshold samples is highlighted using different shades of the same color within each $z$-bin with darker shades representing higher mass thresholds as indicated in the legends. The strength of clustering increases with stellar mass on all scales which is unsurprising because more massive galaxies tend to reside in denser environments. We can also see a break from a power law in the clustering amplitude between small scales and large scales due to the contributions from the pairs within the same halos at small scales (“one-halo”) and from the pairs between different halos (“two-halo”) at large scales as described in Section 4. In each panel, we highlight the scale where the “two-halo” contribution to the real-space correlation function drops below that of “one-halo” with a black marker. This break can be seen to be occurring at different scales in different $z$-bins because of how the comoving transverse distance projected on the same angular scale on the sky changes with redshift due to the changing rate of expansion of the Universe.

The measured SMFs which are jointly fit with clustering to get the HODs are shown in the inset panels via filled circles and error bars. The evolution in SMF with redshift is also highlighted by the lowest $z$-bin SMF overplotted as a dotted magenta curve in all the other panels. The lowest $z$-bin SMF has the characteristic "knee" depicting a sharp drop in the number densities of the high-mass galaxies. This "knee" is barely noticeable, if at all, within the highest $z$-bin but gets more and more pronounced with decreasing redshift, indicative of increasing strength of feedback effects suppressing star formation in high-mass halos/galaxies (J. Silk & M. J. Rees 1998; A. J. Benson et al. 2003). Moreover, the steepening of the low-mass slope with increasing redshift as reported in the literature (I. Davidzon et al. 2017; J. R. Weaver et al. 2023) can be seen. The overplotted lowest $z$-bin SMF in all the other panels also highlights how the dynamic range in stellar mass shrinks with redshift going from over 3 decades in stellar mass at $z \sim 0.35$ (lowest $z$-bin) to only about a decade at $z \sim 4$ (highest $z$-bin). This considerably weakens the SHMR constraining power of the high-$z$ data (clustering + SMF) in isolation. However, the use of "smooth-$z$" modeling in this paper connecting high z with low z using continuity priors remedies this to a large extent as shown in Section 5.2.

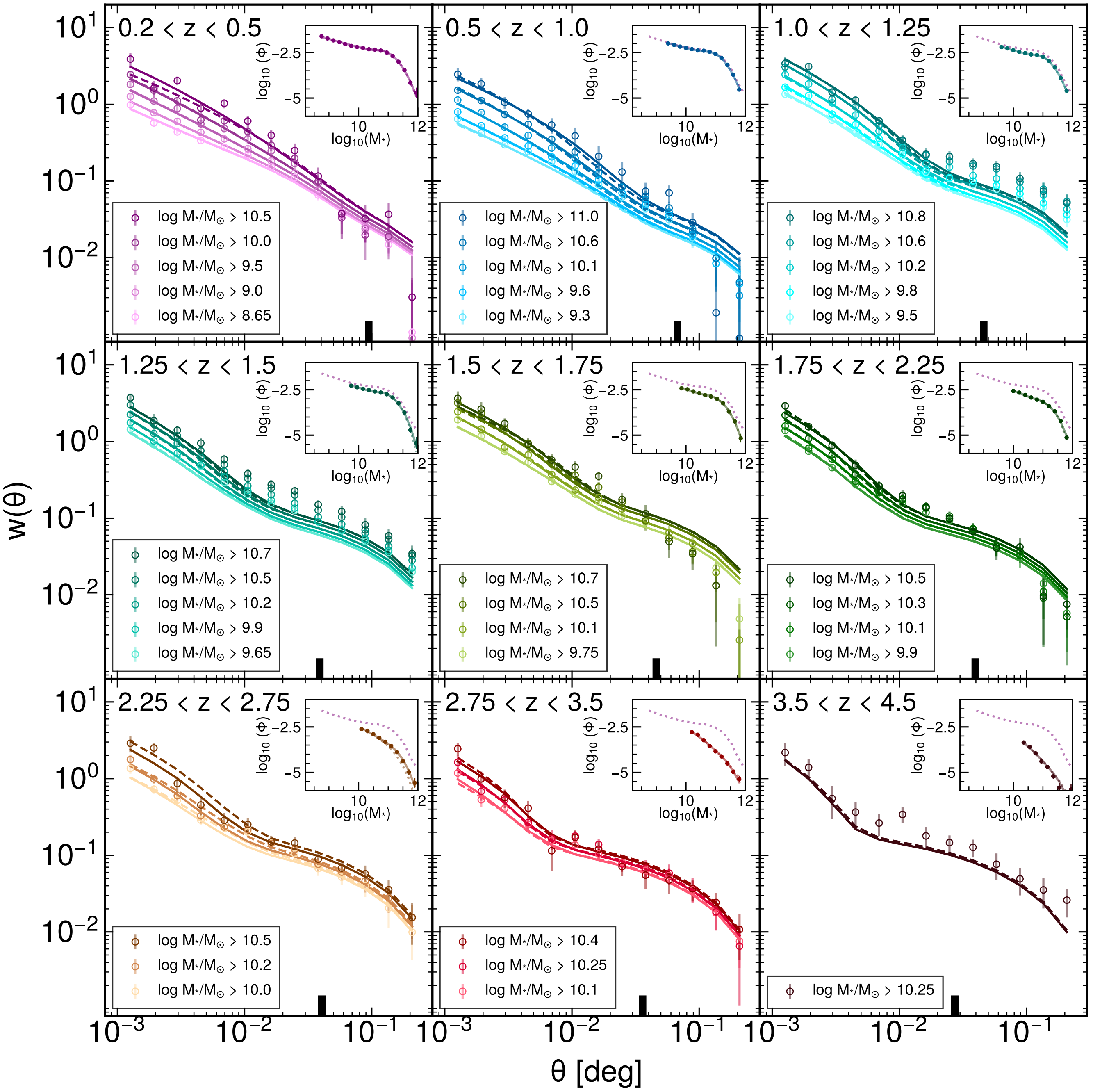

**Figure 4.** Clustering (main panels) and SMF (inset panels) measurements in the whole UDS+COSMOS sample along with the best fits of the "smooth-$z$" model as a function of redshift. The empty circles in the main panels show the clustering measurements and their associated uncertainty, whereas the solid curves show the clustering produced by the best-fit "smooth-$z$" model, our main result, whereas the dashed curves show the results from the "discrete-$z$" model for comparison. Different mass threshold samples are indicated via color shading as indicated in the legends, with darker shades showing higher mass threshold samples. The breaks in scale, where the "two-halo" contribution to $w(\theta)$ overtakes that of "one-halo," are marked in black at the bottom of each panel. In the inset panels, the SMF measurements are shown via filled points and error bars. The lowest redshift bin SMF is also shown as a dotted magenta curve in all the other inset panels to highlight the redshift evolution in the measured SMF. The abundance produced by the best-fit "smooth-$z$" model is shown via solid curves, our main result, and the "discrete-$z$" results are shown via dashed curves for comparison.

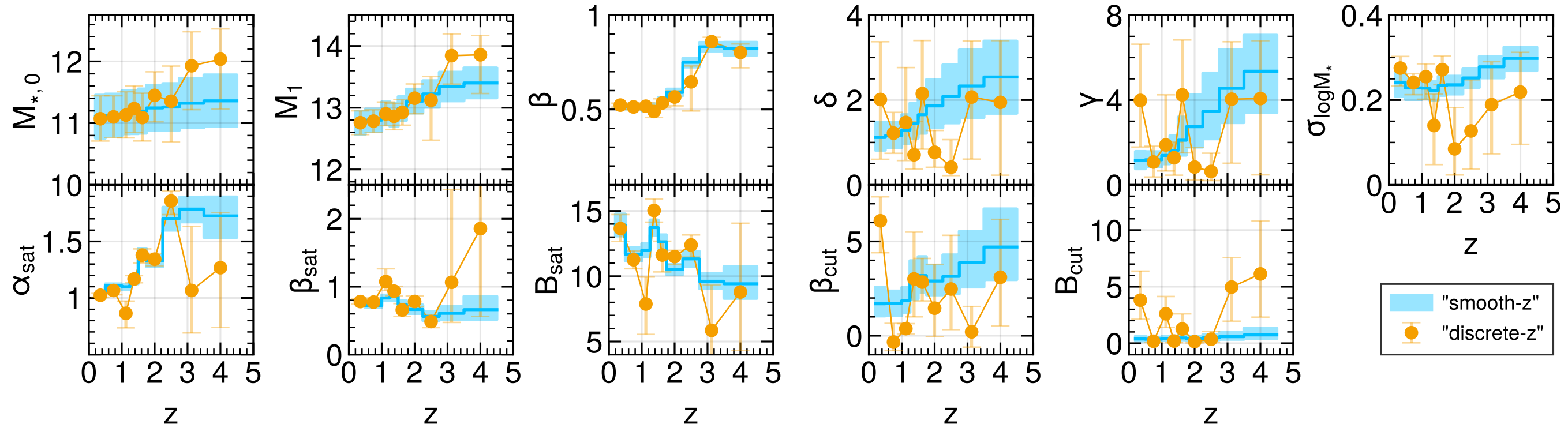

**Figure 5.** HOD parameter fits derived through the "discrete-*z*" model are shown in yellow scatter points and the ones derived using the novel "smooth-*z*" model are shown via the blue line. The errors indicated through yellow bars for "discrete-*z*" and blue-shaded envelope for "smooth-*z*" encompass the posteriors' 16–84 percentile range.

### 5.2. "Discrete-z" versus "Smooth-z" Model Fits

We introduced a novel approach to fitting the HOD model in Section 4.8 wherein we jointly fit clustering and SMF data within all of the nine *z*-bins connected via smooth evolution ("smooth-*z*" model) of the HOD parameters that produce the observed data.

In Figure 5, we show the increased constraining power of the "smooth-*z*" model over the standard "discrete-*z*" model in which the data in the individual *z*-bins are fitted with no regard to the adjacent *z*-bins. For each of the 11 HOD parameters, the yellow scatter points show the medians of the posteriors for the "discrete-*z*" model, whereas the corresponding error bars show the range spanned by their 16th/84th percentiles. As hinted in Section 4.8, with the "discrete-*z*" model, it proved a challenge to constrain the parameters sensitive to the data from the high-stellar-mass end of the SHMR at all redshifts and low-stellar-mass end of the SHMR with increasing redshift.

At the high-mass end, it is the $\delta$ and $\gamma$ that control the rise in the SHMR in the high-mass regime and the transition from low mass to high mass, respectively. Within most *z*-bins, the error bars span a wide range within the bounds on $\delta \in [0.1, 4]$ and $\gamma \in [0, 8]$ meaning that they are very weakly constrained by the data when only considering one *z*-bin at a time. The weak constraining of the high-mass sensitive parameters $\delta$ and $\gamma$ can be attributed to large uncertainties in the SMFs at the high-mass end (and the corresponding contributions to clustering) due to not capturing enough rare massive galaxies. Even though our dataset covers a large effective survey area of $\sim$1.61 deg$^2$, pivotal in capturing a good number of rare massive galaxies, it is still not large enough to capture a large number of rare ultramassive galaxies ($M_* \gtrsim 10^{11}\ M_\odot$ ) to pin down the SMF at the high-mass end with confidence on par with the intermediate stellar masses. Therefore, it makes sense that the HOD parameters sensitive to the high-mass population, $\delta$ and $\gamma$, cannot be constrained better, especially in the "discrete-*z*" model.

At the low-mass end at high *z*, it is challenging to constrain the $\beta$ parameter, which controls the low-mass slope of the SHMR, due to the shrinking dynamical range in complete stellar masses with redshift. Moreover, in five out of nine *z*-bins, the scatter in stellar mass at fixed halo mass, $\sigma_{\log M_*}$, is only constrained within an accuracy of $\sim$0.2 dex which remains inadequate in informing the strength of the stellar mass–halo mass connection across redshifts. A small scatter would mean that the halo mass assembly is tightly linked to the stellar mass assembly in the Universe. In contrast, a large scatter would hint at the stellar mass dependence on other halo properties beyond its mass, a possible indication of galaxy assembly bias. Therefore, it is crucial to constrain $\sigma_{\log M_*}$ better.

The "smooth-*z*" model results are shown via the blue line (median) and the light blue shaded region (16th/84th percentiles) in Figure 5. The three HOD parameters that especially struggled to converge in the "discrete-*z*" model as described above are now remarkably well constrained at all redshifts.

Figure 4 shows that there are subtle differences between the clustering and SMF fits from the "discrete-*z*" (dashed curves) and "smooth-*z*" (solid curves) models. Yet the "smooth-*z*" fits are the result of a continuous redshift evolution of the HOD parameters, which is likely preferred by nature, as opposed to the discontinuous evolution in the "discrete-*z*" model. Quantitatively, the reduced best-fit $\chi^2$ for the "smooth-*z*" fit, $\chi^2_{\rm red, smooth-z} = 2.18$, where the degrees of freedom, DOF $= 472 = 571$ (number of data points within all *z*-bins) - 99 ($11 \times 9$ HOD parameters within all *z*-bins). Within the "discrete-z" model, if we add all the raw best-fit $\chi^2$ from the individual *z*-bin fits, we can then divide the total by the same DOF as in "smooth-*z*", to get $\chi^2_{\rm red, discrete-z}$[14]$= 1.99$, which can be compared to $\chi^2_{\rm red, smooth-z}$ on a similar level.

### 5.3. Redshift Evolution of the mean SHMR for Centrals

The fitted SHMR at each *z*-bin also allows us to paint a continuous picture of the evolution of SFE from $z = 4.5$ to $z = 0.2$. Using the median fitted values of the central HOD parameters in Equation (3) at each redshift, we plot the stellar–halo mass ratios ($M_*/M_h$) as a function of the halo mass in Figure 6, and show the 16th/84th percentiles based on the MCMC run via lightly shaded envelopes. The $M_*/M_h$ of a halo directly indicates how efficient the halo has been, throughout its lifetime, in converting baryonic matter into stellar matter. The SFE, given by dividing $M_*/M_h$ by the baryonic fraction of the universe, $f_b = 15.8\%$ (Planck Collaboration et al. 2020), is indicated on the right in Figure 6.

Furthermore, the lowest and highest halo masses probed within each redshift bin are marked via vertical dashed lines,

[14] the $\chi^2_{\rm red, discrete-z}$ calculated using the median models is actually closer, 2.18, compared to $\chi^2_{\rm red, smooth-z}$= 2.17 calculated using the median models, since some parameters are so poorly constrained in high *z*-bins for "discrete-*z*" that the median is not that close to the best fit

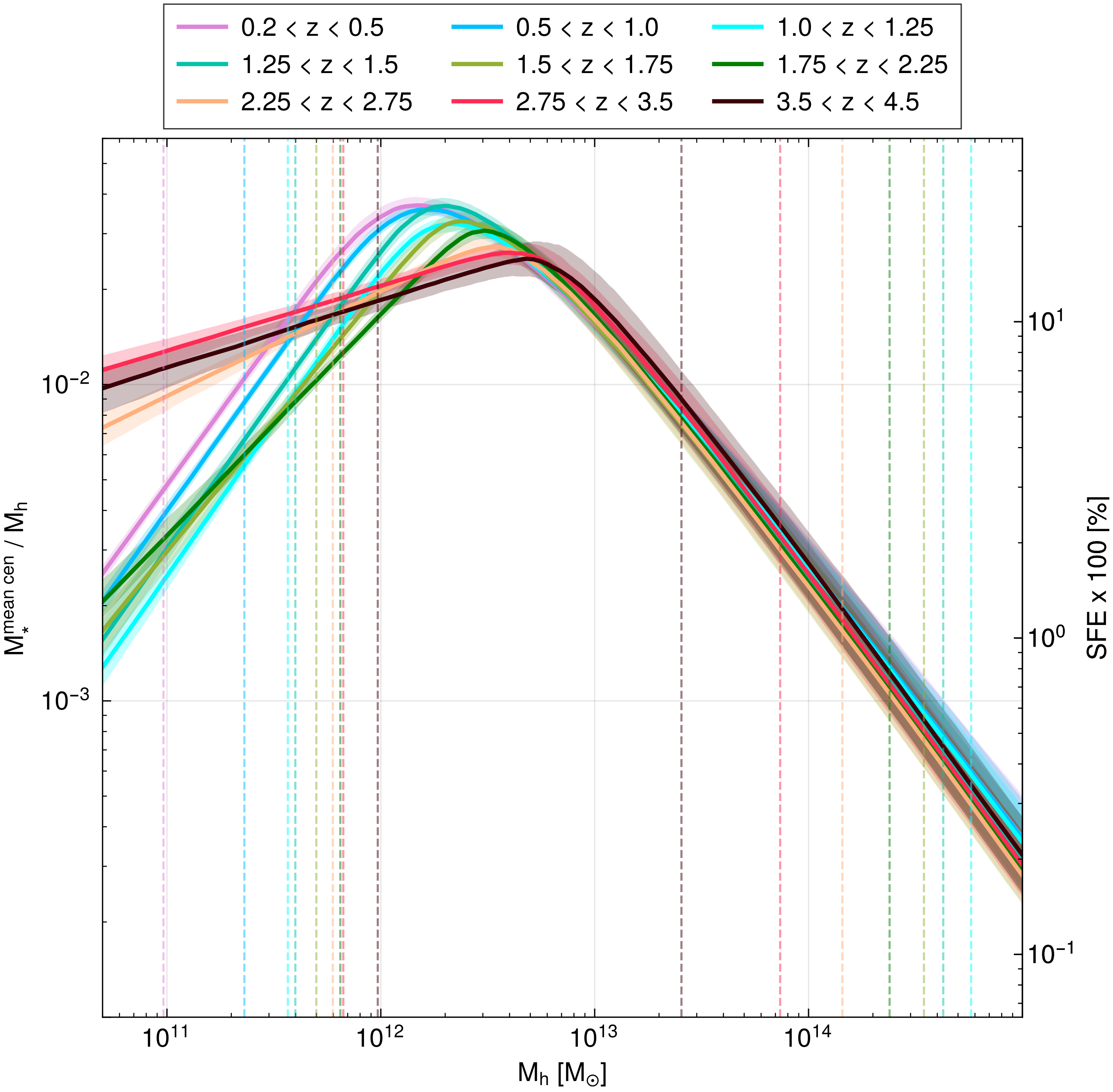

**Figure 6.** Mean stellar–halo mass ratio of central galaxies, $M_{*,\rm meancen}/M_h$ as a function of halo mass for each $z$-bin. The SFE ($= (M_*^{\rm mean\ cen}/M_h)/f_b$) in percentages is indicated on the rightmost ordinates. The colored vertical dashed lines (two for each $z$-bin) indicate the lowest and highest halo mass probed by our analysis, as described in Section 5.3, where the color indicates the $z$-bin as in the legend. Note: the highest halo mass end line (vertical dashed magenta) of the first $z$-bin is missing as it lies at the halo mass beyond the rightmost extent of the $x$-axis.

with color once again indicating redshift as in the legend. The lowest halo mass probed is given by the mean central SHMR at the lowest stellar mass threshold within each $z$-bin. On the other side, the highest halo mass probed is considered where the mean number of galaxies (derived using the fitted HOD at the respective redshift) drops below 0.6 within the volume covered. Even though we extrapolate our models beyond the halo masses directly probed by our analysis, we alert the reader to peruse them with caution.

The low-mass slope starts relatively flat in the highest $z$-bins $z > 2.25$ and steepens the most at $z \sim 2$. This sharp transition at $z \sim 2$ is also evident in the redshift evolution of $\beta$ as seen in Figure 5; higher values indicate flatter slopes. At the low-mass end, smaller-scale feedback effects such as supernovae, stellar winds, etc., are thought to heat up or expel their surrounding gas and suppress star formation (P. F. Hopkins et al. 2012). Therefore, the redshift evolution of the low-mass slope hints at the strengthening of such feedback effects with time.

On the other hand, $M_{*,\rm meancen}/M_h$ declines at the high-mass end at all redshifts. The AGN feedback is purported to be responsible for curbing star formation in massive galaxies (J. Silk & M. J. Rees 1998; A. J. Benson et al. 2003), predominantly in "jet mode" (J. Silk & M. J. Rees 1998; D. J. Croton et al. 2006; R. Kondapally et al. 2023;

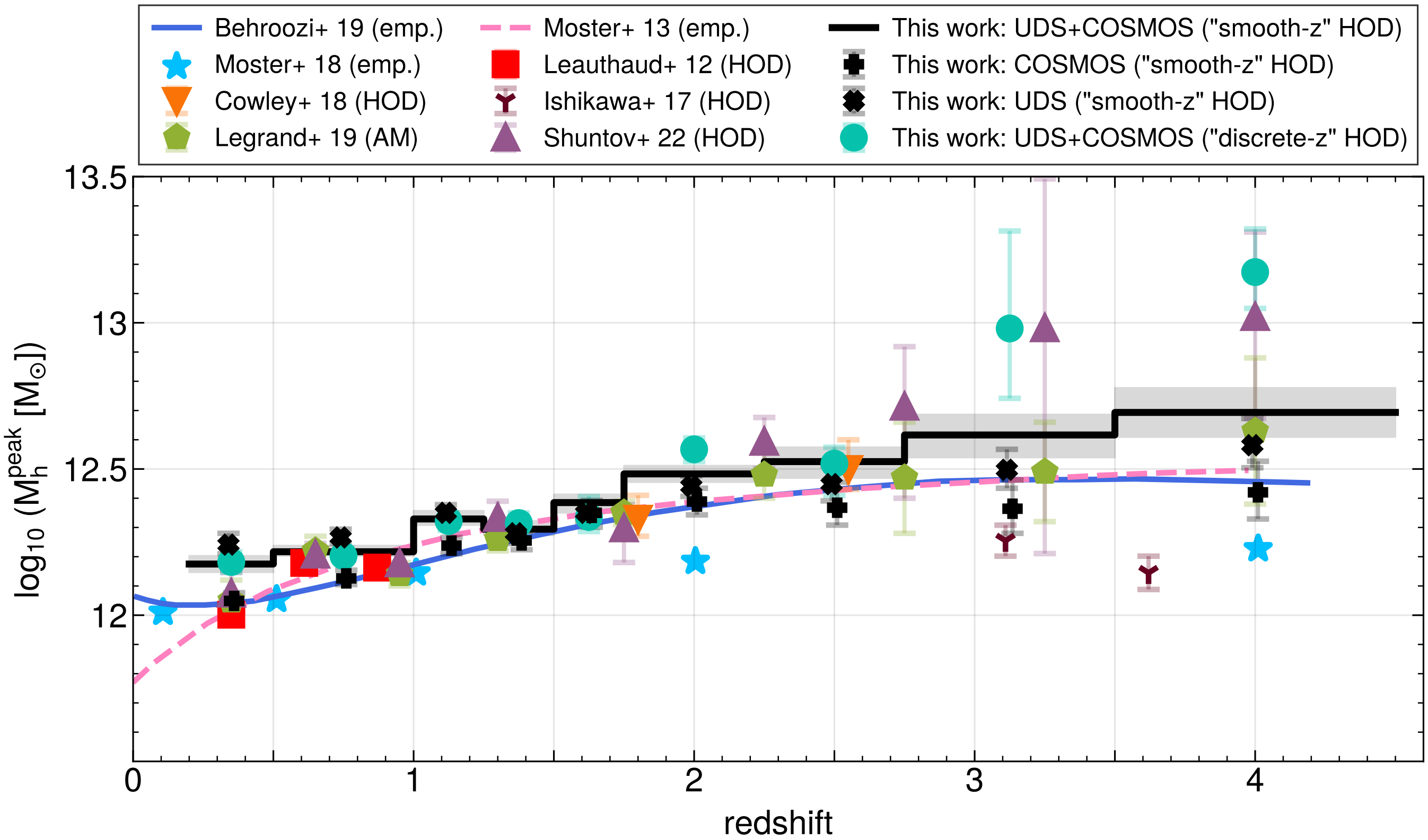

**Figure 7.** Redshift evolution of the halo mass, $M_h^{\rm peak}$, at which the $M_*^{\rm mean\ cen}/M_h$ ratio as shown in Figure 6 peaks, representing the peak SFE (integrated over a halo's lifetime). The solid black line and the gray shaded region shows our main result (using "smooth-$z$") and the corresponding $1\sigma$ uncertainty, respectively, for the full UDS+COSMOS sample. We also indicate the median $M_h^{\rm peak}$ and $1\sigma$ errors using only UDS (black $\times$'s) and COSMOS (black $+$'s) (both with "smooth-$z$"), highlighting the cosmic variance. A compilation of other observationally constrained $M_h^{\rm peak}$ values in the literature using different methods: abundance matching (AM), empirical modeling (emp.), and HOD (equivalent to our "discrete-$z$" HOD model) are also shown. Finally, to highlight the decrease in error on $M_h^{\rm peak}$ going from "discrete-$z$" to "smooth-$z$," especially at $z > 3$, we also include our "discrete-$z$" UDS+COSMOS results as filled turquoise circles.

L. Scharré et al. 2024), explaining the decline in $M_{*,{\rm meancen}}/M_h$ at the high-mass end. Moreover, the high-mass slope remains remarkably consistent throughout the $z$-bins. This indicates that the strength of AGN feedback beyond the peak halo mass is strongly coupled to the halo mass and depends weakly, if at all, on the age of the Universe.

The combination of the low- and high-mass slopes results in a narrow range in halo mass for peak integrated star formation at each redshift, $M_h^{\rm peak} \sim 10^{12.2}$– $10^{12.7}$ $M_\odot$, reflecting an integrated SFE $\sim$16%–23%. This indicates tight coupling between halo mass assembly and galaxy stellar mass, and that quenching strongly depends on halo mass. This tight stellar mass–halo mass connection is further supported by the relatively small scatter in stellar mass at a fixed halo mass, $\sigma_{\log M_*} = 0.22$–$0.30$ dex, across the redshifts probed. The value of $M_h^{\rm peak} \sim 10^{12}$ $M_\odot$ is well supported by a series of observational studies (P. S. Behroozi et al. 2013; P. Behroozi et al. 2019; S. Ishikawa et al. 2017; W. I. Cowley et al. 2018; B. P. Moster et al. 2018; M. Shuntov et al. 2022). It is important to note that even though the adopted L11 model allows for a transition regime between low mass and high mass, it does not necessitate a peak halo mass; it naturally comes out of the fitting.

### 5.4. Redshift Evolution of the Peak Halo Mass, $M_h^{\rm peak}$

To better understand the evolution of the feedback effects suppressing star formation, we now look at the evolution of the halo mass with peak integrated SFE, $M_h^{\rm peak}$, in Figure 7. $M_h^{\rm peak}$ are the peaks in Figure 6. Using the combined sample (UDS +COSMOS), we show the evolution of $M_h^{\rm peak}$ in Figure 7 using a solid black line, with the gray shaded region showing the $1\sigma$ uncertainty evaluated using the MCMC chains. This is our main result. We also show a compilation of results from the literature indicating the method they used to derive $M_h^{\rm peak}$ in the legend, whether it was empirical modeling (emp.), abundance matching (AM), or HOD modeling, like in our case. In our case, there is a downsizing (L. L. Cowie et al. 1996) trend throughout, meaning that the most efficient halo mass ($M_h^{\rm peak}$) declines from $10^{12.69^{+0.08}_{-0.08}}$ $M_\odot$ at $z \sim 4$ to $10^{12.17^{+0.03}_{-0.03}}$ $M_\odot$ in the lowest $z$-bin at $z \sim 0.35$, exhibiting a decline of 0.52 dex. Given the vast range of cosmic history covered, this relatively weak evolution in $M_h^{\rm peak}$ shows that halo mass assembly in halos is tightly linked to the stellar mass assembly.

The relatively small error, $\lesssim +/-0.08$ dex on $M_h^{\rm peak}$, within the highest two $z$-bins ($2.75 < z < 4.5$) can be attributed to the new "smooth-$z$" model (introduced in Section 4.8) used in this study. This model, which connects the HOD parameters across redshift discouraging any sharp inter-$z$-bin "jumps," leverages the high constraining power of the low-$z$ data with a much wider dynamical range in stellar mass than at high $z$. As a result, we could shrink the uncertainty on $\log_{10}(M_h^{\rm peak}$ $[M_\odot])$ at high $z$ by up to $\sim$0.57 dex (maximum improvement occurs at $z \sim 3.2$) compared to M. Shuntov et al. (2022), who applied the same HOD (L11) as we did on a similarly large effective area of $\sim$1.27 deg$^2$, but fitted the individual $z$-bins separately as in our "discrete-$z$" model. For direct comparison, we also show our "discrete-$z$" $M_h^{\rm peak}$ results with filled turquoise circles in Figure 7, where, similar to M. Shuntov et al. (2022), our $M_h^{\rm peak}$ values have much larger uncertainties compared to those from "smooth-$z$" within the highest two $z$-bins ($2.75 < z < 4.5$).

However, for "discrete-$z$" $M_h^{\rm peak}$ (filled turquoise circles) at $z \gtrsim 2$, we note that they are only plotted for completeness and should not be taken at face value as they result from unconstrained HOD parameter posteriors. Therefore, those $M_h^{\rm peak}$ values do not reflect the HOD posteriors where all parameters are constrained (see Figure 14 in Appendix B.1, for instance). Additionally, the "smooth-$z$" $M_h^{\rm peak}$ values for our individual fields at $z > 2$ (black "+" and "x") do not straddle the $M_h^{\rm peak}$ from our UDS+COSMOS (solid black line) as one would intuitively expect. We attribute this to the lack of constraining power of individual fields at the low-mass end, where even the deployment of the "smooth-$z$" model is unable to remediate that. As a result, the posteriors for $\beta$, which controls the low-mass slope of the SHMR, and/or $\alpha_{\rm sat}$, which controls the satellite slope of the HOD, tend to be truncated at the upper end of the flat prior boundaries. However, such is not the case for the posteriors of our main result, UDS+COSMOS "smooth-$z$," even in the highest redshift bin (see Figure 14 in Appendix B.1).

Overall, our results generally agree with the observationally constrained $M_h^{\rm peak}$ in the literature as can be seen in Figure 7. However, at $z < 1.25$, where the wider dynamical range in stellar masses than at higher redshifts provides stronger constraining power for all studies, there seems to be a systematic offset between our $M_h^{\rm peak}$ and others': our $M_h^{\rm peak}$ remains higher by up to ∼0.2 dex at $z < 1.25$. We checked if this discrepancy washes away using the stellar masses derived using `FAST` instead of Dense Basis, but that is not the case: at $z < 0.5$, the use of `FAST` results in a negligible change in $M_h^{\rm peak}$, whereas at $0.5 < z < 1.25$, $M_h^{\rm peak}$ increases further by ∼0.1 dex. Below, we further delve into our investigation of this systematic discrepancy.

#### 5.4.1. Effect of Cosmic Variance on the Redshift Evolution of the Peak Halo Mass at z < 1.25

All of the previous $M_h^{\rm peak}$ results in the literature at these redshifts are based either purely on the COSMOS field (A. Leauthaud et al. 2012; W. I. Cowley et al. 2018; L. Legrand et al. 2019; J. R. Weaver et al. 2022) or rely heavily on the COSMOS field despite having data from other fields in the case of P. Behroozi et al. (2019)[15] and B. P. Moster et al. (2018).

One of the key advantages of this work is the introduction of the UDS as the second field. This difference seems a likely explanation for the offset in our $z < 1.25$ $M_h^{\rm peak}$ results from those in the literature. To test this possibility, we reanalyzed the data from the COSMOS and UDS fields separately to see if our COSMOS-only $M_h^{\rm peak}$ agrees with the literature. If the $M_h^{\rm peak}$ resulting from the clustering and SMF measurements in our COSMOS dataset fitted with the HOD model agreed well with the COSMOS-dominated results in the literature, it would be a clear indication that our UDS+COSMOS $M_h^{\rm peak}$ is driven upward due to the addition of the UDS field. This is exactly what we see in Figure 7, where our COSMOS-only $M_h^{\rm peak}$ results, shown with the black "+," lie almost exactly atop the $M_h^{\rm peak}$ points from other studies. Our UDS-only $M_h^{\rm peak}$ results, shown with the black "×", are higher, which, as speculated, drives the UDS+COSMOS $M_h^{\rm peak}$ upward compared to the COSMOS-only case.

We delve further into the cause of this field-to-field variance through clustering and SMF in Figure 8. In the top row of the figure, we show the clustering offsets for the lowest and highest mass threshold samples within each of our four $z$-bins out to $z = 1.25$. Clustering does not change significantly between the two fields except primarily for the largest scales but the trend is not consistent among the three bins. Therefore, it is hard to infer any resulting change in $M_h^{\rm peak}$.

However, the SMF offset shown in the bottom row of Figure 8 shows that the COSMOS SMFs are systematically overdense (higher normalization) for most of the stellar masses out to $z = 1.25$. Below, we elaborate on the convincing evidence for the generally overdense nature of the COSMOS SMF compared to the UDS SMF:

1. $0.2 < z < 0.5$: At $M_* \lesssim 10^{11.2}\ M_\odot$, the COSMOS SMF offset remains consistently overdense by ∼0.2 dex, and becomes abruptly more overdense at higher masses, reaching an offset of ∼0.55 dex at $M_* \sim 10^{11.8}\ M_\odot$.
2. $0.5 < z < 1.0$: Similar to the previous $z$-bin, at $M_* \lesssim 10^{11.2}\ M_\odot$, the COSMOS SMF remains overdense by ∼0.1 dex, beyond which it abruptly increases in overdensity to reach an offset of ∼0.9 dex at $M_* \sim 10^{11.8} M_\odot$.
3. $1.0 < z < 1.25$: At $M_* \sim 10^{10.2}\ M_\odot$, the COSMOS SMF starts to become overdense steadily until $M_* \sim 10^{11.4}$, before rising sharply to reach an offset of ∼0.5 dex at $M_* \sim 10^{11.8} M_\odot$.

The overdense nature of COSMOS is not surprising as several massive structures, out of which five are $M_* > 10^{13} M_\odot$ have long been reported (N. Scoville et al. 2007b) in the field at $z < 1.1$. There is a superstructure at $z \sim 0.8$–$0.9$ (B. Darvish et al. 2014). D. J. McLeod et al. (2021) have also shown that UDS SMFs are less dense than COSMOS SMFs at $0.25 \leqslant z < 1.25$.

The effect this has on our HOD-modeling-derived $M_h^{\rm peak}$ can be understood by considering the SMFs and the adopted HMF. Due to the higher normalization of COSMOS, more galaxies need to be assigned to halos. The additional galaxies in COSMOS can be accommodated in two ways: (i) be hosted as satellites in high-mass halos, thus increasing the slope of the satellite occupation function, $\alpha_{\rm sat}$, or (ii) by decreasing the halo mass cutoff to host central galaxies, which allows less massive halos to also host galaxies.

According to the first way, the clustering should increase on all scales but we do not see that in Figure 8. Therefore, the second option is more likely—the halo mass cutoff to host central galaxies lowers to accommodate more galaxies as centrals in COSMOS. Inevidently, this lowers the average halo mass essentially captured by the characteristic halo mass, the $M_1$ parameter in the L11 HOD. Since $M_1$ in COSMOS is lower, the effect of which is to shift the SHMR leftwards (toward lower halo masses), the $M_h^{\rm peak}$ also shifts to lower halo masses as we see in Figure 7.

[15] Besides COSMOS-based SMFs, at $0.05 < z < 1$, P. Behroozi et al. (2019) utilize the SMFs from J. Moustakas et al. (2013) which is based on five PRIMUS fields (CDFS, COSMOS, ELAIS-S1, XMM-SXDS, XMM-CFHTLS) covering a large total area of 5.5 deg$^2$. However, due to the photometry problems in J. Moustakas et al. (2013) as discussed in the appendix (C2) of P. Behroozi et al. (2019), they exclude the constraints from J. Moustakas et al. (2013) on the massive end ($M_* > 10^{11}\ M_\odot$) beyond $z > 0.2$. This is also the mass regime where the COSMOS SMF deviates the most from UDS SMF (COSMOS SMF is much overdense) as shown in Figure 8.

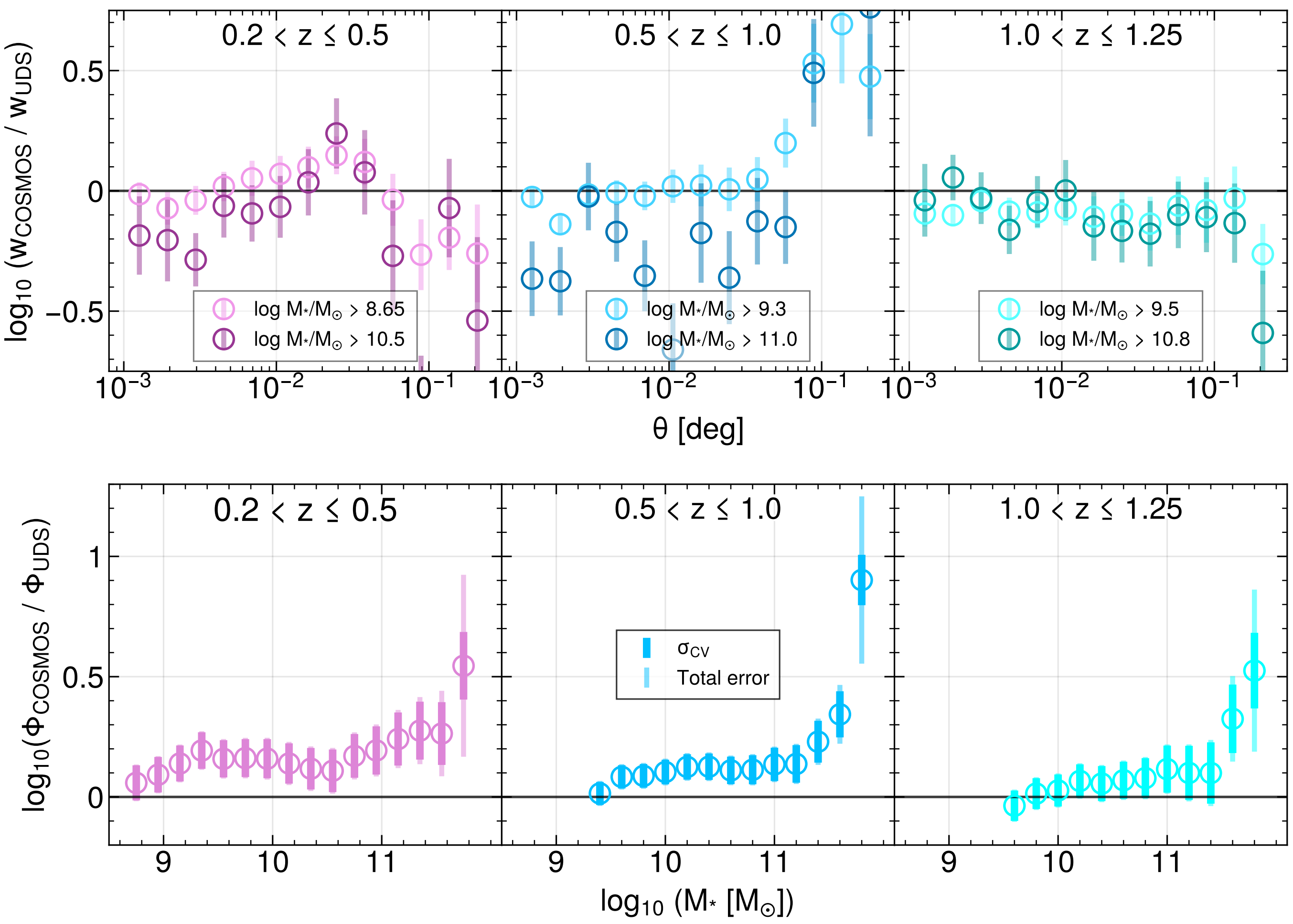

**Figure 8.** Top row: measured clustering offsets between the UDS and COSMOS fields at $z < 1.25$. Each panel shows the offset in the two-point angular correlation functions corrected for integral constraints. For clarity, the offsets are shown only for the lowest and the highest stellar mass threshold samples (as described in Table 1) with the legend indicating the thresholds. Bottom row: measured SMF offsets between the UDS and COSMOS fields $z < 1.25$. Here, we highlight the error budget attributed to cosmic variance ($\sigma_{\rm CV}$) with thicker error bars, whereas the total error including $\sigma_{\rm CV}$, Poisson error ($\sigma_{\rm Poisson}$), and the error due to uncertainty in survey area ($\sigma_{\rm Area}$) are shown with thinner error bars.

### *5.5. Redshift Evolution of the Total (Central + Satellite) SHMR*

Beyond halo masses of about $\sim 10^{12.5}\ M_\odot$, depending on redshift, the contribution of satellites to the total stellar mass budget of the halo starts to become noticeable. The L11 HOD model used in this study allows us to easily calculate the contribution of central and satellites to the total stellar mass locked up in a halo of a given mass through Equation (10), where we essentially integrate the conditional SMF down to a lower limit on stellar mass.

Similar to A. Leauthaud et al. (2012) and M. Shuntov et al. (2022), we integrate down to a stellar mass, $M_* = 10^9\ M_\odot$ and show the resulting total stellar–halo mass ratios, $M_*^{\rm tot}/M_h$ (solid curves) split between centrals (dashed curves) and satellites (dashed–dotted curves) in Figure 9. We also estimate the $1\sigma$ errors on $M_{*,\rm total}/M_h$ using the MCMC samples and show them as shaded regions. Furthermore, as in Figure 6 (and described in Section 5.3), we shade the halo masses not directly probed in our analysis in gray.

Below $M_h \sim 10^{12}\ M_\odot$, the satellite contribution remains subdominant compared to the central galaxy. Beyond $M_h \sim 10^{12}\ M_\odot$, the central galaxies account for a smaller and smaller fraction of the total stellar mass in halos, and the satellites take over beyond $M_h \sim 10^{13.5}\ M_\odot$ (group/cluster-scale halos), with a slight dependence on redshift. Interestingly, the peak total SHMR is dictated almost entirely by centrals and occurs in the satellite subdominant regime as satellites never quite take over centrals in the high-mass regime, within the halo masses directly probed, and occurs at $<M_h^{\rm peak}>$ of $\sim 10^{12.4}\ M_\odot$ (see Section 5.4). This is unsurprising as $M_h^{\rm peak}$ just represents the peak of $M_*^{\rm mean\ cen}/M_h$ and $M_*^{\rm tot}/M_h$ approaches $M_*^{\rm cen}/M_h$ in the satellite subdominant regime as apparent in Figure 9.

Beyond $M_h \sim 10^{13}\ M_\odot$ at $z > 1.5$, there is a hint of an upturn in $M_*^{\rm tot}/M_h$ which could be due to satellites on their way to merge with the central galaxy. At the high-mass end, $M_*^{\rm tot}/M_h$ also rises in the second $z$-bin ($z \sim 0.75$) relative to the third $z$-bin ($z \sim 1.1$), and then falls back again in the lowest $z$-bin ($z \sim 0.35$). This indicates that at $z \sim 0.75$, there is an uptick in the satellite accretion onto the most massive galaxies, and by $z \sim 0.35$ those satellites have mostly merged with the massive central galaxies. We discuss this in Section 6.2 as well.

So far for these total stellar mass calculations we have used a lower stellar mass limit of the integration of $M_* = 10^9 M_\odot$. Ideally, we would like to extend the lower limit down to dwarf galaxies ($10^7$–$10^8$ $M_\odot$) and it is possible to extrapolate our fitted HOD models to such low stellar masses. However, our data are not utilizable below the 80% stellar mass completeness limit, which is already above $10^9$ $M_\odot$ for all but our lowest redshift samples. However, as has previously been noted (A. Leauthaud et al. 2012; M. Shuntov et al. 2022), as long as the low-mass slope of the satellite conditional mass function (see Figure 16 in Appendix C) is sufficiently shallow, lowering the integration limit does not change the satellite contribution significantly.

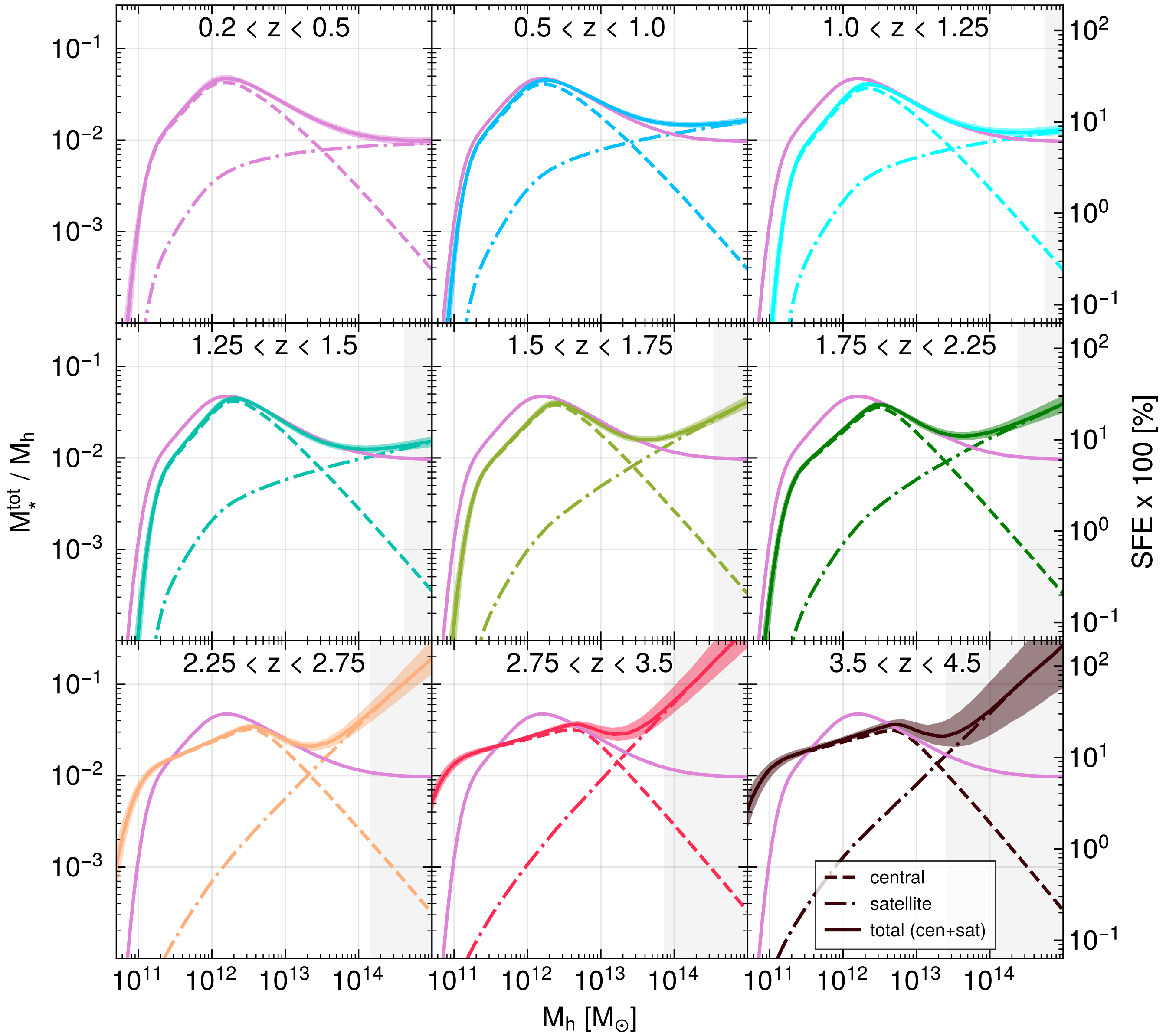

**Figure 9.** Total (central + satellites) stellar mass locked up in halos divided by the halo mass, $M_{*,\rm total}/M_h$ ratio as a function of halo mass within each $z$-bin (each panel). The solid-colored curve shows the median $M_{*,\rm total}/M_h$ and the shaded envelope encapsulates the $1\sigma$ error. The dashed curves show the contribution of central galaxies to the $M_{*,\rm total}/M_h$, whereas the dotted–dashed curves show the contribution from satellite galaxies. The magenta-colored curve showing the $M_{*,\rm total}/M_h$ of the first $z$-bin (top-left panel) is overplotted in all the other panels to highlight its evolution with redshift. As in Figure 6, the rightmost ordinates indicate the SFE ($= (M_*^{\rm tot/cen/sat}/M_h)/f_b$) in percentage. The gray shaded regions indicate the halo masses not directly probed in our analysis, as in Figure 6.

### 5.6. Redshift Evolution of the $M_*^{\rm tot}/M_h$ at Fixed Halo Masses

We present the evolution of $M_*^{\rm tot}/M_h$ (or integrated SFE) in halos of fixed masses—$M_h^{\rm peak}$ ($z$), $10^{11.5}$, $10^{12}$, $10^{12.5}$, and $10^{13.5}$ $M_\odot$ in Figure 10. $M_h^{\rm peak}$ ($z$) halos are always the most efficient by definition which closely trace the $10^{12.5}M_\odot$ halos (in red), indicating that the $M_h^{\rm peak}$ does not deviate much from $\sim10^{12.5}M_\odot$ with redshift, as already described in Section 5.4. At $z < 2$, the slightly lower $M_h = 10^{12}M_\odot$ (in green), and the lowest $M_h = 10^{11.5}M_\odot$ (in blue), continue to rise in integrated SFE with decreasing redshift. However, both generally remain lower in integrated SFE than the $M_h = 10^{12.5}M_\odot$ (or $M_h^{\rm peak}$) halos by factors of $\sim$1.5 and $\sim$4, respectively.

On the other side of $M_h^{\rm peak}$, at $M_h = 10^{13.5}M_\odot$ (in yellow), a reverse trend is seen where the integrated SFE continues to decrease with decreasing redshift (barring a slight uptick at $z \sim 0.75$), indicating that the high-mass halos do bulk of their star formation early on and become more and more quiescent over time.

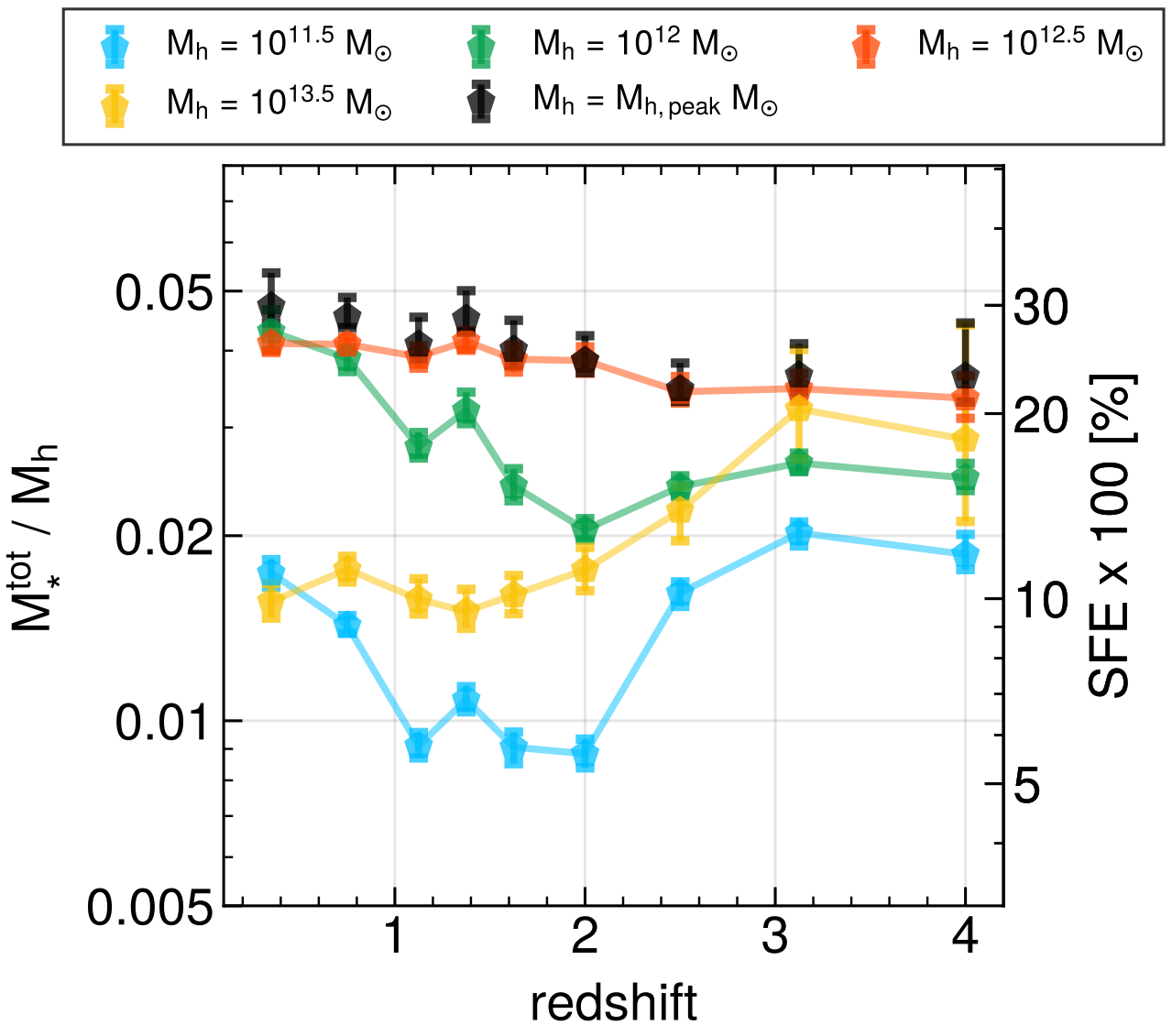

**Figure 10.** Redshift evolution of the total stellar mass-to-halo mass ratio, $M_*^{\rm tot}/M_h$, in halos of fixed mass. Different halo masses are indicated in the legend, and the integrated SFE (= $(M_*^{\rm mean\ cen}/M_h)/f_b$) in percentage is shown on the right $y$-axis.

### *5.7. Redshift Evolution of SHMR Along the Halo Merger Trees*

So far in this paper, halo masses or SHMR (total or otherwise) have been inferred from HOD fitting of large samples of galaxies in fixed $z$-bins. We have then made comparisons of the evolution of various galaxy and halo properties with redshift either at fixed galaxy or halo mass. However, galaxies and halos are both evolving in mass with redshift and so by comparing at fixed mass we are not comparing the "same" halos or galaxies at each redshift. Ideally we would like to trace the halo mass along the merger trees of the same halos. Using our fitted SHMR, we can retrieve the stellar masses of halos along their merger histories from a dark-matter-only simulation. For this analysis, we retrieved the publicly available data from the dark-matter-only TNG300-1-Dark simulation (A. Pillepich et al. 2018). Starting from $z \sim 0.2$ (snapshot number 84), the lowest redshift in our galaxy sample, we trace the subhalo mass history along the main progenitor branch for the 2000 most massive halos. This results in final halo masses at $z \sim 0.2$ ranging from $\sim 10^{12}\ M_\odot$ to $\sim 10^{16}\ M_\odot$ and their halo mass histories.

We divide the final halo masses into bins of 0.2 dex and within each bin trace the mean of the halo mass histories. We convert these mean halo mass histories into $M_*^{\rm tot}/M_h$ histories using the $M_*^{\rm tot}/M_h$ at fixed halo masses shown in Figure 9. By calculating $M_*^{\rm tot}/M_h$ for each mean progenitor halo mass along the merger trees at medians of our $z$-bins, we trace the evolution of $M_*^{\rm tot}/M_h$ as the halo masses grow along the merger trees. The resulting $M_*^{\rm tot}/M_h$ along the merger trees are shown in the left panel of Figure 11. Similarly, we convert the $M_*^{\rm mean\ cen}/M_h$ at fixed halo masses shown in Figure 6 into $M_*^{\rm mean\ cen}/M_h$ along the merger trees and show the results in the right panel of Figure 11. In Figure 11, we indicate the mean halo masses with the sizes of the circles in the legend at the top. The final mean log halo mass, $\log_{10}$ ($M_h$ [$M_\odot$]), and the mean central galaxy stellar masses, $\log_{10}$ ($<M_{*\,\rm cen}>$ [$M_\odot$]) at $z = 0.35$ (median of the lowest $z$-bin), are highlighted in different shades of red, as shown in the legend at the bottom, to guide the eye to trace their evolution.

Perhaps the most visually striking aspect of Figure 11 is the change in the hierarchy of integrated SFE in halos with redshift. The highest mass halos have the highest integrated SFE at $z \gtrsim 3$ and the order gets completely flipped by $z = 0.35$, where the lowest mass halos have the highest integrated SFE. For instance, the highest average mass of halos ($M_h = 10^{13.8}\ M_\odot$) at $z = 0.35$ shown in the darkest shade of red accumulates most of its stellar mass early on at $z \gtrsim 3$, both in terms of the total stellar mass and that of only centrals, with the integrated SFEs peaking at $\sim$22% and $\sim$18%, respectively. At $z \lesssim 3$, this representative halo is found to be continually declining in integrated SFE, getting completely quenched by $z = 0.35$.

This is in stark contrast to the lowest mean halo mass ($M_h = 10^{12.27}\ M_\odot$) considered at $z = 0.35$ which started building its stellar content much more recently at $z \lesssim 2$, with its integrated SFE peaking ($\sim$30% for the total stellar mass buildup in the left panel) only at $z \sim 0.35$. This is a clear sign of downsizing (L. L. Cowie et al. 1996) manifesting in the decline in the halo mass scale over the time at which the integrated SFE peaks—the highest mass halos assemble their stellar mass first followed by lesser and lesser massive halos.

Comparing the evolution of $M_*^{\rm mean\ cen}/M_h$ (right panel) with $M_*^{\rm tot}/M_h$ (left panel), the increasing contribution of satellites with decreasing redshift is apparent and unsurprising as demonstrated by the overall increase in satellite fraction toward low redshift as seen in Figure 17. It also shows how the stellar mass growth in the most massive halos at low redshift is being driven almost entirely by satellite galaxies, with the opposite true of the lowest mass halos.

## 6. Discussion

The evolution of SHMR has imprints of various physical processes integrated over halos' lifetimes such as cold gas flows toward the centers of halos giving rise to star formation, stellar/dark matter halo mass growth due to mergers, and, on the flip side, the feedback effects that heat up or eject the gas out of galaxies quenching star formation. In light of our SHMR results in Section 5, here we try to build an intuition of the galaxy–halo connection and in turn its connection to galaxy formation.

### *6.1. The Most Star-forming Halo Mass—*$M_h^{\rm peak}$

We shall first look at the redshift evolution of the halo mass with peak integrated SFE, $M_h^{\rm peak}$, as shown in Figure 7 (black line). This peak mass where the central galaxies peak in integrated SFE equally applies to the total SHMR where the peak is dominated by the centrals.

#### *6.1.1. Downsizing in* $M_h^{\rm peak}$

Interestingly, $M_h^{\rm peak}$ drops by 0.52 dex from $z \sim 4$ to $z \sim 0.35$, as described in Section 5.4. Previous $M_h^{\rm peak}$ results in the literature (shown in Figure 7) had seen similar downsizing trends, and notably much stronger than ours in M. Shuntov et al. (2022) at $z > 1.5$. However, since these earlier results had been heavily based on a single field of COSMOS, it was not clear if the downsizing trends in $M_h^{\rm peak}$ were commonplace in the Universe. The use of two large fields in our work (UDS in conjunction with COSMOS) greatly mitigated the effect of cosmic variance on $M_h^{\rm peak}$ evolution, and in doing so we still find a clearly persisting downsizing signal throughout the redshifts probed.

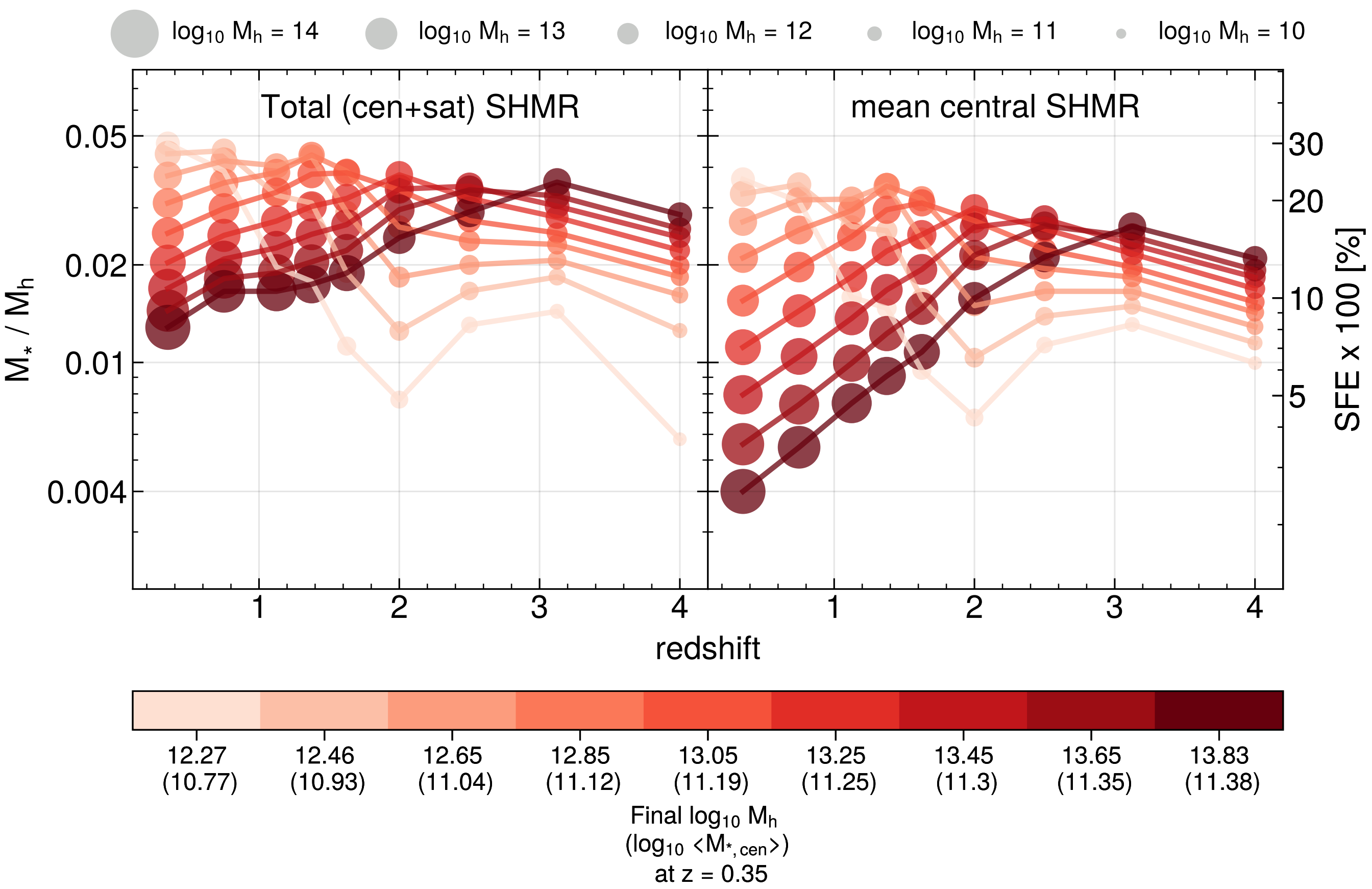

**Figure 11.** Left: evolution of the total (centrals + satellites) stellar–halo mass ratio along the main progenitor branches (most massive branch) of $z = 0.35$ (median of the lowest $z$-bin in our analysis) dark matter halos in the TNG300-Dark-1 simulation (A. Pillepich et al. 2018). The total stellar mass in a given TNG300-Dark-1 halo mass (average in a 0.2 dex bin), depicted by the sizes of the red circles, is calculated using the fitted HOD model parameters at each redshift as described in Section 5.5. The shading of the circles and the lines connecting them reflect the final halo mass at $z = 0.35$ (as shown in the legend), where lighter shades of red depict lower final halo masses. Right: evolution of the mean central stellar–halo mass ratio along the main progenitor branches of $z = 0.35$ dark matter halos in the TNG300-Dark-1 simulation as in the left panel. Again, this is evaluated at the best-fit HOD parameters. As in the left panel, the colors of the circles represent the final halo masses (at $z = 0.35$), and their sizes capture the current halo masses.

#### *6.1.2. Implications on the Strength of AGN Feedback*

The $M_h^{\rm peak}$ observed at any redshift results from the SFH that preceded in a way to result in the most stellar mass per halo mass ($M_*/M_h$). Therefore, however minimal, the change in the $M_h^{\rm peak}$ holds clues to the strength of star formation regulating AGN feedback (J. Silk & M. J. Rees 1998; A. J. Benson et al. 2003) especially at the high-mass end. AGN feedback supposedly acts on the high-mass halos ($\gtrsim 10^{12} M_\odot$) at least in the redshift range considered in this study. There are two broad classes of AGNs in the literature: (i) "jet-mode" AGN, which expel their energy primarily through mechanical jets as they are radiatively inefficient and have low optical/ultraviolet (UV)/X-ray luminosities and low accretion rates with a thick toroidal disk, and generally host more massive black holes and are found more in redder, more massive quiescent galaxies and richer environments; (ii) "adiative-mode" AGN, which are quite the opposite as their radiatively efficient thin disks allow them to expel their energy primarily through radiation instead of jets and are therefore much more luminous in optical/UV/X-ray wavelengths (S. Cielo et al. 2018; T. M. Heckman & P. N. Best 2014).

There is accumulating observational evidence that "jet-mode" AGN are responsible for much of the quenching in massive galaxies. For instance, R. Kondapally et al. (2023) analyzed a sample of ~1000 radio-excess AGNs within $0.5 < z < 2.5$ from the radio catalogs of LoTSS Deep Fields DR1 (J. Sabater et al. 2021; C. Tasse et al. 2021) and found that low-excitation (jet-mode) AGNs dominate the feedback across the full redshift range considered. On the contrary, "radiative-mode" AGNs are actually found to coexist with high levels of star formation. J. Florez et al. (2020) compared a sample of X-ray-luminous AGNs with the galaxies that did not host an X-ray-luminous AGN within $0.5 < z < 3$. They found X-ray-luminous AGNs to be hosting much higher (by a factor of ~3−10) star formation rates across the full range of redshifts and stellar mass considered. J. Florez et al. (2021) find these results to be in general agreement with the simulations SAG (S. A. Cora et al. 2018) and IllustriTNG (R. Weinberger et al. 2017; A. Pillepich et al. 2018), while they see the opposite in their comparison to SIMBA (R. Davé et al. 2019), i.e., in SIMBA, high-X-ray-luminous AGNs are in fact found to have much lower levels of star formation and many are even quenched.

Therefore, for the following, we assume that it is the "jet-mode" AGN feedback which is the dominant suppressor of star formation in massive galaxies and simply call it AGN feedback. Moving from high redshift to low redshift, if the AGN feedback had been weaker recently, higher-mass halos become the most efficient in star formation, shifting $M_h^{\rm peak}$ toward higher masses. On the other hand, stronger AGN feedback in the recent past of the observed redshift manifests in a lower $M_h^{\rm peak}$ at the observed redshift than otherwise.

Given the decrease in $M_h^{\rm peak}$ with time, it is likely that the AGN feedback progresses by quenching the massive galaxies in decreasing order of stellar mass. This is also supported by the hierarchy of quenching as seen in Figure 11, where the most massive galaxies quench first followed by lesser and lesser massive galaxies. By $z \sim 0.5$, most of the massive galaxies in massive halos have quenched, whereas the lower-mass halos are now peaking in integrated SFE, thereby shifting the $M_h^{\rm peak}$ to its lowest value at $z \sim 0.35$.

### 6.2. Stellar–Halo Mass Growth

Figure 11 beautifully captures the downsizing trend in stellar–halo mass growth, i.e., they grow in decreasing order of mass. Group- and cluster-scales halos ($M_h \gtrsim 10^{13.5}$) at $z \sim 0.3$ hosting a quenched galaxy population accumulated the bulk of the stellar mass early on, at $z \gtrsim 2$. Below $z \lesssim 2$, they are found with residual levels of star formation. On the other hand, the less massive halos ($M_h \lesssim 10^{12.7}\ M_\odot$) at $z \sim 0.3$ only started to build up their stellar mass at $z \lesssim 2$, reaching peak levels of star formation at $z \sim 0.3$. Now, we will examine the stellar–halo mass growth in these two regimes more closely.

#### 6.2.1. Low-mass Regime

Throughout the redshift range considered in this study ($0.2 < z < 4.5$), on the lower side of the peak stellar–halo mass ratio (either of the $M_h^{\rm peak}$ or the peak in $M_*^{\rm tot}/M_h$), stellar mass increases with increasing halo mass. As time passes, halos are expected to grow via merging with much smaller halos with lower stellar mass (A. Leauthaud et al. 2012; L. Legrand et al. 2019). Therefore, with an increase in halo mass, there will be a mild increase in stellar mass, but the $M_*^{\rm tot}/M_h$ should decline. Mergers alone cannot explain the steep rise in $M_*^{\rm tot}/M_h$ at $z \lesssim 2$; secular "in situ" star formation has to be invoked. At $z \gtrsim 2$, the low-mass slope becomes shallower indicating lower levels of "in situ" star formation which is not surprising as star-forming efficiencies of low-mass halos only pick up at $z \lesssim 2$, as seen in Figure 11.

#### 6.2.2. High-mass Regime

As visible in Figure 9, the central galaxies dominate stellar mass within halos below the peak at $M_h \sim 10^{12}\ M_\odot$. Beyond this mass, in group-scale halos, the central contribution starts to decline and is taken over by the satellites. At first glance, this might indicate that the central galaxies suffer from a decline in integrated SFE because the star-forming ability of the massive halos ($M_h \gtrsim 10^{12}\ M_\odot$) now gets distributed between centrals and satellites. However, upon close inspection, this does not seem to be the case, at least at $z \lesssim 1.5$. In the "satellite-dominated" regime (beyond $M_h \sim 10^{12}\ M_\odot$), even if all the stellar mass in satellites is somehow transferred to centrals, i.e., the solid $M_*^{\rm tot}/M_h$ curve becomes the $M_*^{\rm cen}/M_h$ curve, the peak integrated SFE remains at $M_h \sim 10^{12}\ M_\odot$, i.e., there is no second peak on the high-mass end. This means that centrals in group-scale halos do not exceed the integrated SFE obtained at $M_h \sim 10^{12}\ M_\odot$, not because the satellites take away their share of integrated SFE, but rather that beyond $M_h \sim 10^{12}\ M_\odot$, there is a global decline in integrated SFE of halos, again at least out to $z \sim 1.5$. Whatever quenching mechanism is at play in group-scale halos impacts both the central and satellites. This is in general agreement with A. Leauthaud et al. (2012) and M. Shuntov et al. (2022).

Interestingly at $z > 1.5$, $M_*^{\rm tot}/M_h$ seems to rise again beyond $M_h \sim 10^{13}\ M_\odot$ but it is hard to ascertain if it indeed overtakes the peak $M_*^{\rm tot}/M_h$ ratio obtained at $M_h \sim 10^{12}\ M_\odot$. This is because the upper limit on the halo mass probed by our analysis does not extend to high enough halo masses disallowing us to confidently conclude if $M_*^{\rm tot}/M_h$ indeed overtakes the peak of $M_h \sim 10^{12}\ M_\odot$. Even if we extrapolate our models beyond the maximum halo mass probed at each redshift, the highest physically possible halo mass also decreases with increasing redshift, thus leaving a very faint possibility of $M_*^{\rm tot}/M_h$ having a second peak at higher halo masses ($M_h \gtrsim 10^{13}\ M_\odot$).

At $z > 1.5$, regardless if there is a second peak in $M_h$ or not, there is an upturn in $M_*^{\rm tot}/M_h$ beyond $M_h \sim 10^{13}\ M_\odot$ due to satellites. The presence of this upturn and the lack thereof at $z < 1.5$ possibly conveys an important message: at $z > 1.5$, satellite accretion onto the massive halos is more prevalent, which is plausible as the Universe was indeed more merger ridden back then (F. Governato et al. 1999; K. R. Stewart et al. 2009; A. R. Wetzel et al. 2009). However, at $z < 1.5$, many of these satellites might have merged onto central galaxies, hence the suppression in the satellite contribution to $M_*^{\rm tot}/M_h$ for high halo masses.

## 7. Summary

We present a detailed analysis of the SHMR of central and satellite galaxies since $z = 4.5$. To derive the SHMRs, we fit the HOD models using measurements of galaxy clustering and abundance in two large legacy UDS and COSMOS fields, covering a total effective area of ∼1.61 deg$^2$. This is the first time that two photometric datasets, FENIKS v2 (UDS; K. Zaidi et al. 2024) and UVISTA DR3 ultradeep stripes (COSMOS; A. Muzzin et al. 2013), of such area coverage and depth have been homogeneously integrated to trace stellar mass assembly within halos.

We also introduce a novel HOD fitting framework that allows the high constraining power of low-$z$ data, due to more than ∼3 dex of dynamical range in complete stellar masses, to communicate with high $z$ ($z \gtrsim 2.5$), where we are complete only for ≲1 dex. Data in all $z$-bins are jointly fitted where HOD parameters between adjacent $z$-bins are connected via continuity (smoothing) priors. This is motivated by the assumption that galaxy clustering and abundance measurements between adjacent $z$-bins evolve from one to another, instead of being independent of each other. Therefore, HOD parameters between neighboring $z$-bins should not see abrupt changes. This allowed dramatically reducing the uncertainties in SHMR at $z > 2.5$, giving a much clearer picture of SHMR evolution out to $z = 4.5$. Our findings are as follows:

1. We find that the evolution of halo mass with peak integrated SFE, $M_h^{\rm peak}$, is consistent with downsizing found in COSMOS-only or COSMOS-dominated studies. $M_h^{\rm peak}$ declines by ∼0.5 dex, decreasing from $10^{12.69^{+0.08}_{-0.08}}\ M_\odot$ at $z \sim 4$ to $10^{12.17^{+0.03}_{-0.03}}\ M_\odot$ at $z \sim 0.35$.
2. At $z < 1.25$, using our UDS+COSMOS sample, we observe systematically high $M_h^{\rm peak}$ by ∼0.2 dex than found by many COSMOS-only or COSMOS-dominated studies (see Section 5.4). Upon investigation, we find that this discrepancy is indeed linked to cosmic variance:

overdense SMFs of COSMOS due to many reported overdense structures (see Section 5.4.1), in the context of HOD modeling, decrease the low halo mass cutoff to host galaxies. This manifests in shifting our COSMOS-only $M_h^{\rm peak}$ to systematically lower values, consistent with the COSMOS-only or COSMOS-dominated values reported in the literature.
3. The total stellar mass locked up in halos less massive than $M_h \sim 10^{12} M_\odot$ is dominated by central galaxies, whereas the satellites dominate the stellar mass budget beyond $M_h \sim 10^{13} M_\odot$ in group- and cluster-scale halos (see Section 5.5).
4. We show in Figure 11 the evolution in integrated SFE of halos, for the first time in the context of HOD modeling, as they grow along their progenitor merger trees. Star formation occurs hierarchically: most massive halos and galaxies observed within the lowest $z$-bin ($z \sim 0.35$) accumulate their stellar mass first followed by progressively lighter systems. This hierarchy applies to both central and satellite galaxies (see Section 5.7).
5. The conditional SMF of central galaxies depends strongly on halo mass. It peaks at increasingly larger stellar masses for increasing halo masses, and tails off on either end. Satellites' conditional SMFs, however, depend much more weakly on halo mass; satellites are equally scattered across $M_h = 10^{11.5}$–$10^{13.5} M_\odot$ at $M_* \lesssim 10^{10}\ M_\odot$. Overall, the conditional satellite SMFs have a classic SMF shape with a power law at low masses and a "knee" at intermediate masses, beyond which it declines sharply (see Appendix C).
6. The fraction of mass in satellites, $f_{\rm sat}$, generally increases with decreasing redshift and stellar mass thresholds. $f_{\rm sat}$ never exceeds ∼28% attained within the lowest $z$-bin and stellar mass threshold considered (see Appendix D).

## Acknowledgments

K.Z. acknowledges the John F. Burlingamme Graduate Fellowship in the Physics & Astronomy Department at Tufts University for support while conducting majority of this work. K.Z. and D.M. acknowledge support from the National Science Foundation under grant AST-2009442. Additionally, C.P. and J.A.D. acknowledge support from the National Science Foundation under grant AST-2009632. K.G. acknowledges support from Australian Research Council Laureate Fellowship FL180100060.

The authors acknowledge the Tufts University High Performance Compute Cluster (https://it.tufts.edu/high-performance-computing) which was utilized for the research reported in this paper.

*Software*: `Astropy`(Astropy Collaboration et al. 2022), `numpy`(C. R. Harris et al. 2020), `halomod`(S. G. Murray et al. 2021), `CORRFUNC` (M. Sinha & L. H. Garrison 2020; M. Sinha & L. Garrison 2019), `pycorr` (https://github.com/cosmodesi/pycorr).

## Appendix A
## Measured Stellar Mass Functions

In addition to the comparison between the SMFs of UDS (FENIKS v2 catalog; K. Zaidi et al. 2024) and COSMOS (ultradeep stripes of UVISTA DR3; A. Muzzin et al. 2013) fields in Section 5.4.1 at $z < 1.25$, we present the raw SMFs for both the individual fields and their combination in Figure 12. The joint use of two, instead of the usual one, homogeneously prepared photometric catalogs covering large sky areas is one of the novelties of this paper, key for mitigating errors due to cosmic variance, $\sigma_{\rm CV}$. We highlight the contribution of $\sigma_{\rm CV}$ to the total error budget for the SMFs in Figure 12 through thick bars while the total error is shown via thinner bars. The decrease in $\sigma_{\rm CV}$ can be easily seen through the diminishing extent of these thick bars in the third panel as compared to their extent in the first two panels showing the SMFs of the individual fields.

In Tables 3–11, we also present the combined SMFs (UDS +COSMOS), including the percentage errors due to the Poisson ($\sigma_{\rm Poisson}$), cosmic variance ($\sigma_{\rm CV}$), and the fixed (3%) uncertainty in the survey area. Note that since $\sigma_{Poisson}$ is not always symmetric, its upper and lower values are quoted with a preceding "+/−," while the $\sigma_{\rm CV}$ and $\sigma_{\rm Area}$ are symmetric.

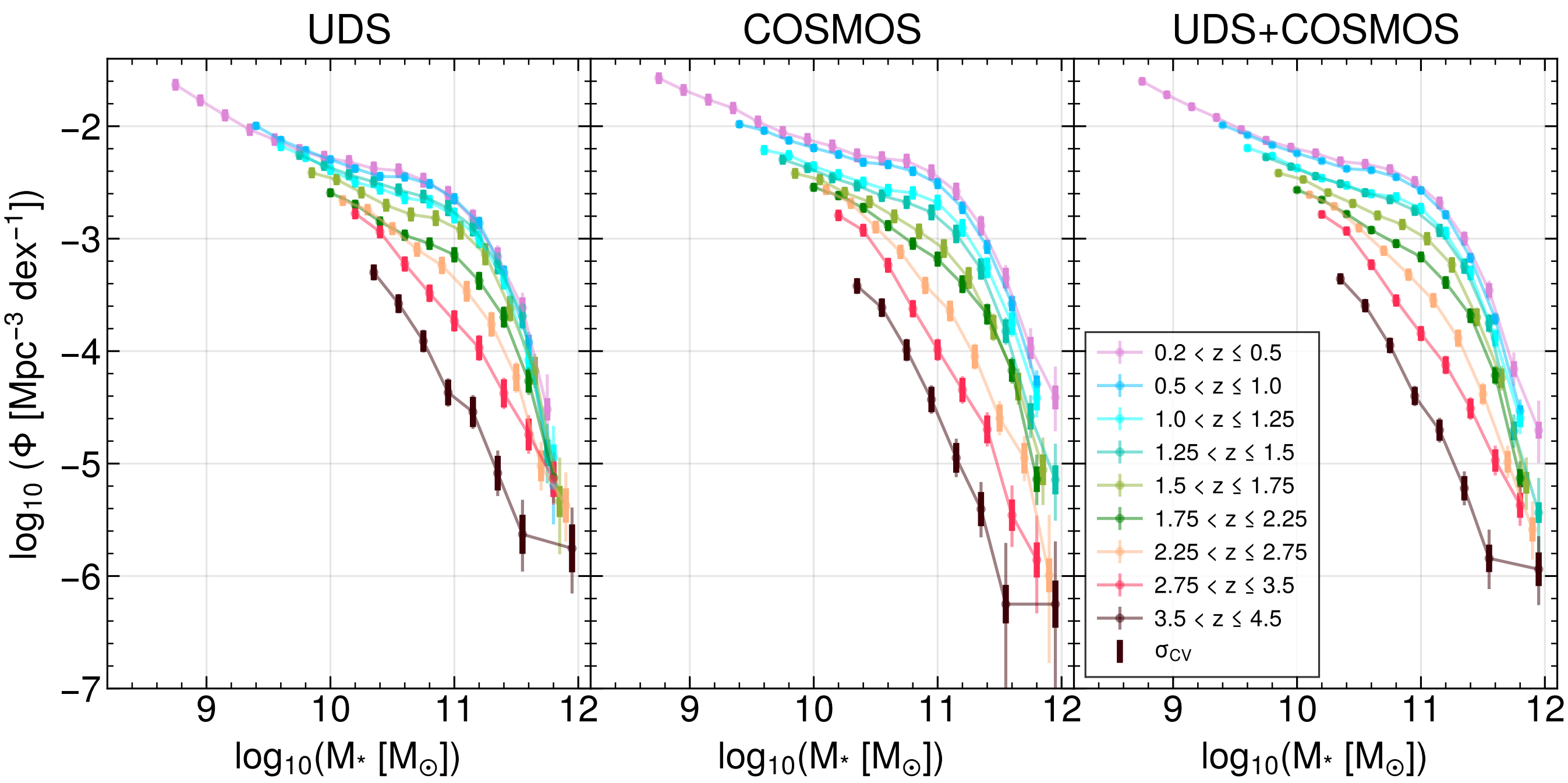

**Figure 12.** Measured SMFs from the individual UDS and COSMOS fields and their combination (UDS+COSMOS). The thinner error bars indicate the total errors as described in Section 3.3, whereas the contributions of cosmic variance (indicated via thicker bars) decrease due to combining the two fields in the rightmost panel.

**Table 3**
$0.2 < z < 0.5$

| $\log_{10}(M_* \ [M_\odot])_{\rm bin}$ | $N_{\rm gal}$ | $\log_{10}(\Phi \ [{\rm Mpc}^{-3} \ {\rm dex}^{-1}])$ | $\sigma_{\rm Poisson}$ | $\sigma_{\rm CV}$ | $\sigma_{\rm Area}$ |
|---|---|---|---|---|---|
| 8.65–8.85 | 5065 | $-1.6007^{+0.0392}_{-0.0392}$ | $^{+1.41\%}_{-1.42\%}$ | 8.4% | 3% |
| 8.85–9.05 | 3844 | $-1.7205^{+0.0398}_{-0.0398}$ | $^{+1.63\%}_{-1.63\%}$ | 8.51% | 3% |
| 9.05–9.25 | 3010 | $-1.8267^{+0.0404}_{-0.0404}$ | $^{+1.84\%}_{-1.84\%}$ | 8.62% | 3% |
| 9.25–9.45 | 2416 | $-1.9221^{+0.0412}_{-0.0412}$ | $^{+2.05\%}_{-2.06\%}$ | 8.75% | 3% |
| 9.45–9.65 | 1881 | $-2.0309^{+0.0421}_{-0.0421}$ | $^{+2.33\%}_{-2.33\%}$ | 8.92% | 3% |
| 9.65–9.85 | 1509 | $-2.1266^{+0.043}_{-0.043}$ | $^{+2.61\%}_{-2.61\%}$ | 9.08% | 3% |
| 9.85–10.05 | 1306 | $-2.1893^{+0.0444}_{-0.0444}$ | $^{+2.8\%}_{-2.81\%}$ | 9.36% | 3% |
| 10.05–10.25 | 1161 | $-2.2404^{+0.0458}_{-0.0458}$ | $^{+2.98\%}_{-2.98\%}$ | 9.66% | 3% |
| 10.25–10.45 | 983 | $-2.3127^{+0.0472}_{-0.0473}$ | $^{+3.24\%}_{-3.24\%}$ | 9.94% | 3% |
| 10.45–10.65 | 932 | $-2.3358^{+0.0484}_{-0.0484}$ | $^{+3.33\%}_{-3.33\%}$ | 10.21% | 3% |
| 10.65–10.85 | 833 | $-2.3846^{+0.0497}_{-0.0497}$ | $^{+3.52\%}_{-3.53\%}$ | 10.47% | 3% |
| 10.85–11.05 | 647 | $-2.4943^{+0.0546}_{-0.0546}$ | $^{+4.01\%}_{-4.01\%}$ | 11.54% | 3% |
| 11.05–11.25 | 420 | $-2.682^{+0.0607}_{-0.0607}$ | $^{+4.99\%}_{-5.0\%}$ | 12.7% | 3% |
| 11.25–11.45 | 202 | $-2.9999^{+0.0692}_{-0.0693}$ | $^{+7.27\%}_{-7.29\%}$ | 13.86% | 3% |
| 11.45–11.65 | 70 | $-3.4601^{+0.0862}_{-0.0864}$ | $^{+12.61\%}_{-12.69\%}$ | 15.03% | 3% |
| 11.65–11.85 | 14 | $-4.1591^{+0.1472}_{-0.151}$ | $^{+29.61\%}_{-30.62\%}$ | 16.19% | 3% |
| 11.85–12.05 | 4 | $-4.7032^{+0.2638}_{-0.2919}$ | $^{+58.14\%}_{-64.87\%}$ | 17.35% | 3% |

**Table 4**
$0.5 < z < 1.0$

| $\log_{10}(M_*\ [M_\odot])_{\rm bin}$ | $N_{\rm gal}$ | $\log_{10}(\Phi\ [{\rm Mpc}^{-3}\ {\rm dex}^{-1}])$ | $\sigma_{\rm Poisson}$ | $\sigma_{\rm CV}$ | $\sigma_{\rm Area}$ |
|---|---|---|---|---|---|
| 9.3–9.5 | 9804 | $-1.9892^{+0.0277}_{-0.0277}$ | $^{+1.02\%}_{-1.02\%}$ | 5.52% | 3% |
| 9.5–9.7 | 7984 | $-2.0784^{+0.028}_{-0.028}$ | $^{+1.13\%}_{-1.13\%}$ | 5.59% | 3% |
| 9.7–9.9 | 6507 | $-2.1672^{+0.0284}_{-0.0284}$ | $^{+1.25\%}_{-1.25\%}$ | 5.68% | 3% |
| 9.9–10.1 | 5495 | $-2.2406^{+0.0293}_{-0.0293}$ | $^{+1.36\%}_{-1.36\%}$ | 5.89% | 3% |
| 10.1–10.3 | 4704 | $-2.3081^{+0.0302}_{-0.0302}$ | $^{+1.47\%}_{-1.47\%}$ | 6.09% | 3% |
| 10.3–10.5 | 4005 | $-2.378^{+0.0312}_{-0.0312}$ | $^{+1.59\%}_{-1.59\%}$ | 6.33% | 3% |
| 10.5–10.7 | 3889 | $-2.3908^{+0.0322}_{-0.0322}$ | $^{+1.62\%}_{-1.62\%}$ | 6.59% | 3% |
| 10.7–10.9 | 3380 | $-2.4517^{+0.0338}_{-0.0338}$ | $^{+1.73\%}_{-1.73\%}$ | 6.96% | 3% |
| 10.9–11.1 | 2570 | $-2.5707^{+0.038}_{-0.038}$ | $^{+1.99\%}_{-1.99\%}$ | 7.98% | 3% |
| 11.1–11.3 | 1562 | $-2.7869^{+0.0427}_{-0.0427}$ | $^{+2.56\%}_{-2.56\%}$ | 9.0% | 3% |
| 11.3–11.5 | 646 | $-3.1704^{+0.0487}_{-0.0487}$ | $^{+4.01\%}_{-4.01\%}$ | 10.02% | 3% |
| 11.5–11.7 | 185 | $-3.7134^{+0.0597}_{-0.0597}$ | $^{+7.61\%}_{-7.63\%}$ | 11.04% | 3% |
| 11.7–11.9 | 28 | $-4.5335^{+0.1039}_{-0.1052}$ | $^{+20.44\%}_{-20.79\%}$ | 12.06% | 3% |

**Table 5**
$1.0 < z < 1.25$

| $\log_{10}(M_*\ [M_\odot])_{\rm bin}$ | $N_{\rm gal}$ | $\log_{10}(\Phi\ [{\rm Mpc}^{-3}\ {\rm dex}^{-1}])$ | $\sigma_{\rm Poisson}$ | $\sigma_{\rm CV}$ | $\sigma_{\rm Area}$ |
|---|---|---|---|---|---|
| 9.5–9.7 | 4570 | $-2.1942^{+0.0348}_{-0.0348}$ | $^{+1.49\%}_{-1.49\%}$ | 7.27% | 3% |
| 9.7–9.9 | 3887 | $-2.2645^{+0.0353}_{-0.0353}$ | $^{+1.62\%}_{-1.62\%}$ | 7.39% | 3% |
| 9.9–10.1 | 3034 | $-2.3721^{+0.0367}_{-0.0367}$ | $^{+1.83\%}_{-1.83\%}$ | 7.67% | 3% |
| 10.1–10.3 | 2468 | $-2.4618^{+0.038}_{-0.038}$ | $^{+2.03\%}_{-2.03\%}$ | 7.96% | 3% |
| 10.3–10.5 | 2126 | $-2.5266^{+0.0398}_{-0.0398}$ | $^{+2.19\%}_{-2.19\%}$ | 8.38% | 3% |
| 10.5–10.7 | 1794 | $-2.6003^{+0.042}_{-0.042}$ | $^{+2.39\%}_{-2.39\%}$ | 8.88% | 3% |
| 10.7–10.9 | 1677 | $-2.6296^{+0.0448}_{-0.0448}$ | $^{+2.47\%}_{-2.47\%}$ | 9.56% | 3% |
| 10.9–11.1 | 1326 | $-2.7316^{+0.0521}_{-0.0521}$ | $^{+2.78\%}_{-2.78\%}$ | 11.27% | 3% |
| 11.1–11.3 | 801 | $-2.9505^{+0.0599}_{-0.0599}$ | $^{+3.59\%}_{-3.6\%}$ | 12.97% | 3% |
| 11.3–11.5 | 354 | $-3.3051^{+0.0692}_{-0.0693}$ | $^{+5.45\%}_{-5.46\%}$ | 14.68% | 3% |
| 11.5–11.7 | 93 | $-3.8856^{+0.0864}_{-0.0865}$ | $^{+10.87\%}_{-10.92\%}$ | 16.39% | 3% |
| 11.7–11.9 | 18 | $-4.5988^{+0.1378}_{-0.1402}$ | $^{+25.88\%}_{-26.56\%}$ | 18.1% | 3% |

**Table 6**
$1.25 < z < 1.5$

| $\log_{10}(M_*\ [M_\odot])_{\rm bin}$ | $N_{\rm gal}$ | $\log_{10}(\Phi\ [{\rm Mpc}^{-3}\ {\rm dex}^{-1}])$ | $\sigma_{\rm Poisson}$ | $\sigma_{\rm CV}$ | $\sigma_{\rm Area}$ |
|---|---|---|---|---|---|
| 9.65–9.85 | 4352 | $-2.2746^{+0.0352}_{-0.0352}$ | $^{+1.53\%}_{-1.53\%}$ | 7.38% | 3% |
| 9.85–10.05 | 3580 | $-2.3594^{+0.0365}_{-0.0365}$ | $^{+1.69\%}_{-1.69\%}$ | 7.66% | 3% |
| 10.05–10.25 | 2945 | $-2.4442^{+0.0378}_{-0.0378}$ | $^{+1.86\%}_{-1.86\%}$ | 7.95% | 3% |
| 10.25–10.45 | 2521 | $-2.5117^{+0.0396}_{-0.0396}$ | $^{+2.01\%}_{-2.01\%}$ | 8.38% | 3% |
| 10.45–10.65 | 2099 | $-2.5913^{+0.0423}_{-0.0423}$ | $^{+2.21\%}_{-2.21\%}$ | 9.0% | 3% |
| 10.65–10.85 | 1831 | $-2.6506^{+0.045}_{-0.045}$ | $^{+2.36\%}_{-2.36\%}$ | 9.63% | 3% |
| 10.85–11.05 | 1466 | $-2.7472^{+0.0526}_{-0.0527}$ | $^{+2.65\%}_{-2.65\%}$ | 11.44% | 3% |
| 11.05–11.25 | 969 | $-2.927^{+0.0613}_{-0.0613}$ | $^{+3.26\%}_{-3.26\%}$ | 13.39% | 3% |
| 11.25–11.45 | 466 | $-3.2449^{+0.0709}_{-0.071}$ | $^{+4.74\%}_{-4.74\%}$ | 15.34% | 3% |
| 11.45–11.65 | 145 | $-3.7519^{+0.0849}_{-0.085}$ | $^{+8.63\%}_{-8.66\%}$ | 17.29% | 3% |
| 11.65–11.85 | 16 | $-4.7092^{+0.1466}_{-0.1495}$ | $^{+27.57\%}_{-28.38\%}$ | 19.24% | 3% |
| 11.85–12.05 | 3 | $-5.4362^{+0.3087}_{-0.3522}$ | $^{+67.78\%}_{-78.21\%}$ | 21.2% | 3% |

**Table 7**
$1.5 < z < 1.75$

| $\log_{10}(M_*\ [M_\odot])_{\rm bin}$ | $N_{\rm gal}$ | $\log_{10}(\Phi\ [{\rm Mpc}^{-3}\ {\rm dex}^{-1}])$ | $\sigma_{\rm Poisson}$ | $\sigma_{\rm CV}$ | $\sigma_{\rm Area}$ |
|---|---|---|---|---|---|
| 9.75–9.95 | 3405 | $-2.4169^{+0.037}_{-0.037}$ | $^{+1.73\%}_{-1.73\%}$ | 7.79% | 3% |
| 9.95–10.15 | 2988 | $-2.4736^{+0.0384}_{-0.0384}$ | $^{+1.85\%}_{-1.85\%}$ | 8.1% | 3% |
| 10.15–10.35 | 2283 | $-2.5905^{+0.0399}_{-0.0399}$ | $^{+2.11\%}_{-2.11\%}$ | 8.41% | 3% |
| 10.35–10.55 | 1827 | $-2.6872^{+0.043}_{-0.043}$ | $^{+2.37\%}_{-2.37\%}$ | 9.14% | 3% |
| 10.55–10.75 | 1431 | $-2.7933^{+0.0465}_{-0.0465}$ | $^{+2.68\%}_{-2.68\%}$ | 9.92% | 3% |
| 10.75–10.95 | 1189 | $-2.8738^{+0.0523}_{-0.0523}$ | $^{+2.94\%}_{-2.94\%}$ | 11.28% | 3% |
| 10.95–11.15 | 888 | $-3.0006^{+0.0619}_{-0.0619}$ | $^{+3.41\%}_{-3.41\%}$ | 13.5% | 3% |
| 11.15–11.35 | 519 | $-3.2338^{+0.0722}_{-0.0722}$ | $^{+4.48\%}_{-4.49\%}$ | 15.73% | 3% |
| 11.35–11.55 | 178 | $-3.6986^{+0.0859}_{-0.086}$ | $^{+7.76\%}_{-7.78\%}$ | 17.95% | 3% |
| 11.55–11.75 | 51 | $-4.2414^{+0.1097}_{-0.11}$ | $^{+14.89\%}_{-15.03\%}$ | 20.18% | 3% |
| 11.75–11.95 | 6 | $-5.1708^{+0.2255}_{-0.2398}$ | $^{+46.75\%}_{-50.38\%}$ | 22.41% | 3% |

**Table 8**
$1.75 < z < 2.25$

| $\log_{10}(M_*\ [M_\odot])_{\rm bin}$ | $N_{\rm gal}$ | $\log_{10}(\Phi\ [{\rm Mpc}^{-3}\ {\rm dex}^{-1}])$ | $\sigma_{\rm Poisson}$ | $\sigma_{\rm CV}$ | $\sigma_{\rm Area}$ |
|---|---|---|---|---|---|
| 9.9–10.1 | 5125 | $-2.5664^{+0.0311}_{-0.0311}$ | $^{+1.41\%}_{-1.41\%}$ | 6.35% | 3% |
| 10.1–10.3 | 4194 | $-2.6535^{+0.0322}_{-0.0322}$ | $^{+1.56\%}_{-1.56\%}$ | 6.59% | 3% |
| 10.3–10.5 | 3126 | $-2.7811^{+0.0347}_{-0.0347}$ | $^{+1.8\%}_{-1.8\%}$ | 7.18% | 3% |
| 10.5–10.7 | 2252 | $-2.9235^{+0.038}_{-0.038}$ | $^{+2.13\%}_{-2.13\%}$ | 7.94% | 3% |
| 10.7–10.9 | 1706 | $-3.0441^{+0.0421}_{-0.0421}$ | $^{+2.45\%}_{-2.45\%}$ | 8.89% | 3% |
| 10.9–11.1 | 1299 | $-3.1625^{+0.0503}_{-0.0503}$ | $^{+2.81\%}_{-2.81\%}$ | 10.83% | 3% |
| 11.1–11.3 | 769 | $-3.3902^{+0.0592}_{-0.0592}$ | $^{+3.67\%}_{-3.67\%}$ | 12.77% | 3% |
| 11.3–11.5 | 390 | $-3.685^{+0.069}_{-0.069}$ | $^{+5.19\%}_{-5.19\%}$ | 14.71% | 3% |
| 11.5–11.7 | 115 | $-4.2154^{+0.0848}_{-0.0849}$ | $^{+9.73\%}_{-9.77\%}$ | 16.65% | 3% |
| 11.7–11.9 | 14 | $-5.13^{+0.1524}_{-0.1561}$ | $^{+29.61\%}_{-30.62\%}$ | 18.59% | 3% |

**Table 9**
$2.25 < z < 2.75$

| $\log_{10}(M_*\ [M_\odot])_{\rm bin}$ | $N_{\rm gal}$ | $\log_{10}(\Phi\ [{\rm Mpc}^{-3}\ {\rm dex}^{-1}])$ | $\sigma_{\rm Poisson}$ | $\sigma_{\rm CV}$ | $\sigma_{\rm Area}$ |
|---|---|---|---|---|---|
| 10.0–10.2 | 4740 | $-2.6093^{+0.0364}_{-0.0364}$ | $^{+1.46\%}_{-1.46\%}$ | 7.7% | 3% |
| 10.2–10.4 | 3739 | $-2.7123^{+0.0382}_{-0.0382}$ | $^{+1.65\%}_{-1.65\%}$ | 8.11% | 3% |
| 10.4–10.6 | 2405 | $-2.904^{+0.0431}_{-0.0431}$ | $^{+2.06\%}_{-2.06\%}$ | 9.24% | 3% |
| 10.6–10.8 | 1506 | $-3.1073^{+0.0482}_{-0.0482}$ | $^{+2.61\%}_{-2.61\%}$ | 10.37% | 3% |
| 10.8–11.0 | 923 | $-3.3199^{+0.0571}_{-0.0571}$ | $^{+3.34\%}_{-3.35\%}$ | 12.36% | 3% |
| 11.0–11.2 | 545 | $-3.5487^{+0.0684}_{-0.0684}$ | $^{+4.37\%}_{-4.38\%}$ | 14.83% | 3% |
| 11.2–11.4 | 251 | $-3.8854^{+0.0813}_{-0.0813}$ | $^{+6.5\%}_{-6.51\%}$ | 17.3% | 3% |
| 11.4–11.6 | 80 | $-4.382^{+0.1007}_{-0.1009}$ | $^{+11.76\%}_{-11.83\%}$ | 19.76% | 3% |
| 11.6–11.8 | 20 | $-4.9841^{+0.1441}_{-0.146}$ | $^{+24.46\%}_{-25.04\%}$ | 22.23% | 3% |
| 11.8–12.0 | 5 | $-5.5861^{+0.2487}_{-0.2676}$ | $^{+51.58\%}_{-56.37\%}$ | 24.7% | 3% |

**Table 10**
$2.75 < z < 3.5$

| $\log_{10}(M_*\ [M_\odot])_{\rm bin}$ | $N_{\rm gal}$ | $\log_{10}(\Phi\ [{\rm Mpc}^{-3}\ {\rm dex}^{-1}])$ | $\sigma_{\rm Poisson}$ | $\sigma_{\rm CV}$ | $\sigma_{\rm Area}$ |
|---|---|---|---|---|---|
| 10.1–10.3 | 4627 | $-2.7845^{+0.0388}_{-0.0388}$ | $^{+1.48\%}_{-1.48\%}$ | 8.28% | 3% |
| 10.3–10.5 | 3294 | $-2.9321^{+0.0432}_{-0.0432}$ | $^{+1.76\%}_{-1.76\%}$ | 9.31% | 3% |
| 10.5–10.7 | 1654 | $-3.2312^{+0.0496}_{-0.0496}$ | $^{+2.49\%}_{-2.49\%}$ | 10.74% | 3% |
| 10.7–10.9 | 794 | $-3.55^{+0.0574}_{-0.0574}$ | $^{+3.61\%}_{-3.61\%}$ | 12.35% | 3% |
| 10.9–11.1 | 404 | $-3.8434^{+0.07}_{-0.07}$ | $^{+5.09\%}_{-5.1\%}$ | 15.0% | 3% |
| 11.1–11.3 | 212 | $-4.1234^{+0.0837}_{-0.0837}$ | $^{+7.09\%}_{-7.11\%}$ | 17.65% | 3% |
| 11.3–11.5 | 87 | $-4.5103^{+0.1017}_{-0.1018}$ | $^{+11.25\%}_{-11.31\%}$ | 20.31% | 3% |
| 11.5–11.7 | 30 | $-4.9727^{+0.132}_{-0.1329}$ | $^{+19.71\%}_{-20.02\%}$ | 22.96% | 3% |
| 11.7–11.9 | 12 | $-5.3706^{+0.1791}_{-0.1834}$ | $^{+32.17\%}_{-33.44\%}$ | 25.61% | 3% |

**Table 11**
$3.5 < z < 4.5$

| $\log_{10}(M_*\ [M_\odot])_{\rm bin}$ | $N_{\rm gal}$ | $\log_{10}(\Phi\ [{\rm Mpc}^{-3}\ {\rm dex}^{-1}])$ | $\sigma_{\rm Poisson}$ | $\sigma_{\rm CV}$ | $\sigma_{\rm Area}$ |
|---|---|---|---|---|---|
| 10.25–10.45 | 1529 | $-3.3573^{+0.0519}_{-0.0519}$ | $^{+2.59\%}_{-2.59\%}$ | 11.28% | 3% |
| 10.45–10.65 | 883 | $-3.5957^{+0.0614}_{-0.0614}$ | $^{+3.42\%}_{-3.42\%}$ | 13.38% | 3% |
| 10.65–10.85 | 391 | $-3.9495^{+0.0721}_{-0.0721}$ | $^{+5.18\%}_{-5.19\%}$ | 15.48% | 3% |
| 10.85–11.05 | 139 | $-4.3987^{+0.0901}_{-0.0902}$ | $^{+8.82\%}_{-8.85\%}$ | 18.55% | 3% |
| 11.05–11.25 | 69 | $-4.7028^{+0.1101}_{-0.1103}$ | $^{+12.7\%}_{-12.79\%}$ | 21.73% | 3% |
| 11.25–11.45 | 21 | $-5.2195^{+0.1503}_{-0.1519}$ | $^{+23.83\%}_{-24.37\%}$ | 24.9% | 3% |
| 11.45–11.65 | 5 | $-5.8427^{+0.2554}_{-0.2738}$ | $^{+51.58\%}_{-56.37\%}$ | 28.08% | 3% |
| 11.85–12.05 | 4 | $-5.9396^{+0.2938}_{-0.3192}$ | $^{+58.14\%}_{-64.87\%}$ | 34.44% | 3% |

## Appendix B
## "Discrete-$z$" versus "Smooth-$z$"

### B.1. Example HOD Posteriors for "smooth-z" versus "Discrete-z" Models

We showcase the improvement in fitting due to the implementation of the "smooth-$z$" model over the "discrete-$z$" model, especially at high $z$, through their corner plots showing HOD posteriors in Figures 13–14. At all redshifts, but especially at high redshift (see Figure 14), we see a stark improvement with "smooth-$z$" in how the HOD parameters are constrained.

### B.2. Comparison of the mean SHMR for Centrals between "Smooth-z" and "Discrete-z" Models

In Figure 15, we compare the $M_*^{\rm mean\ \ cen}/M_h$ resulting from the "smooth-$z$" model fits in blue with the "discrete-$z$" model fits in orange, highlighting the gain in precision, which generally becomes more noticeable with increasing redshift and/or increasing halo mass, by employing the "smooth-$z$" model. The disagreement between the two results, which is more pronounced at the high-mass end, could be attributed to the noisy clustering and number density measurements of the rare high-mass galaxies. This couples with weakening constraining power of a singular redshift bin with increasing redshift due to the increasing stellar mass completeness limit. These two limitations of the data on the low and high-mass ends can also be visualized through the narrowing range in the halo masses probed with increasing redshift, indicated via the enlarging gray shaded regions showing the unprobed halo mass windows. As the "discrete-$z$" fits rely on singular redshift bins, as opposed to the "smooth-$z$" fits which compensate for the shrinking dynamical range in mass with redshift via jointly fitting all redshift bins assuming a smooth redshift evolution of the HOD parameters, we are able to constrain the $M_*^{\rm mean\ cen}/M_h$ out to our highest $z$-bin with much more certainty.

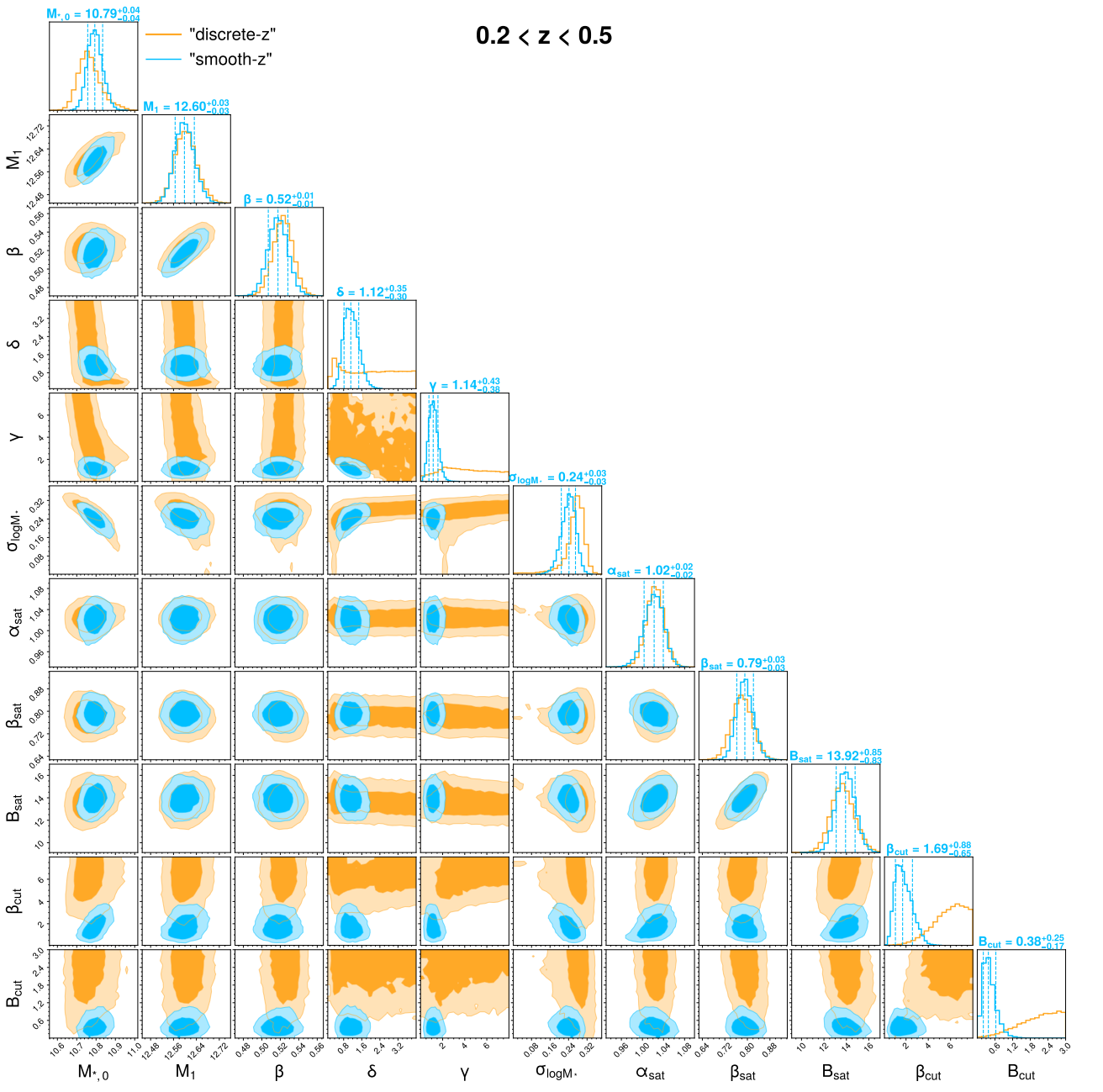

**Figure 13.** Corner plot showing HOD posteriors at $0.2 < z < 0.5$ from "smooth-$z$" fits (in blue), where we jointly fitted data in all the $z$-bins out to $z = 4.5$ (as described in Section 4.8), and "discrete-$z$" fits (in yellow). The median values and the $1\sigma$ uncertainty of "smooth-$z$" fits are shown on top of the 1D projected marginalized distributions along the diagonal. The 2D projected marginalized $1\sigma$ and $2\sigma$ distributions are shown with blue and yellow contours, again for "smooth-$z$" fits and "discrete-$z$" fits, respectively.

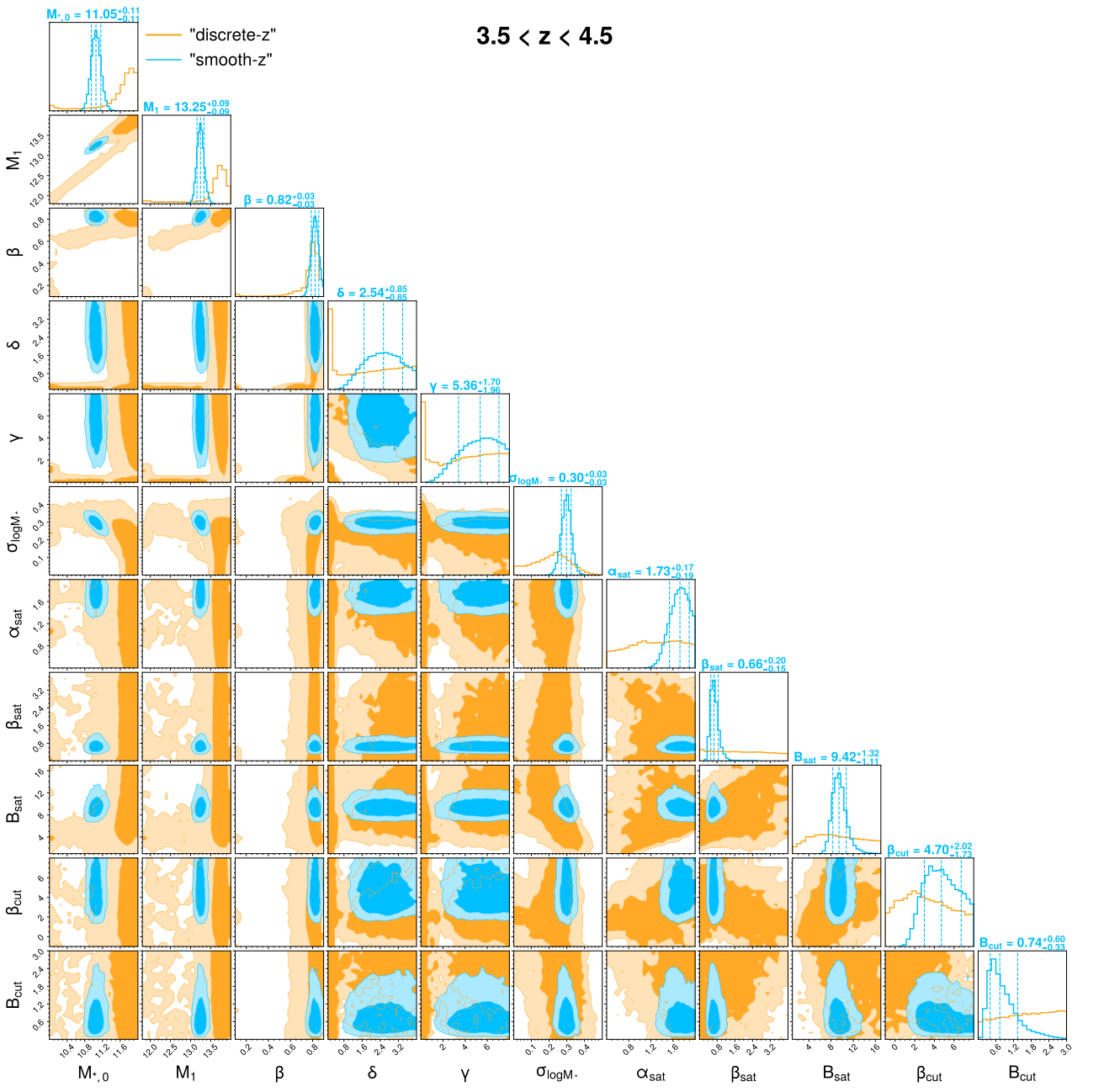

**Figure 14.** Same as Figure 13 but at $3.5 < z < 4.5$ (the highest $z$-bin).

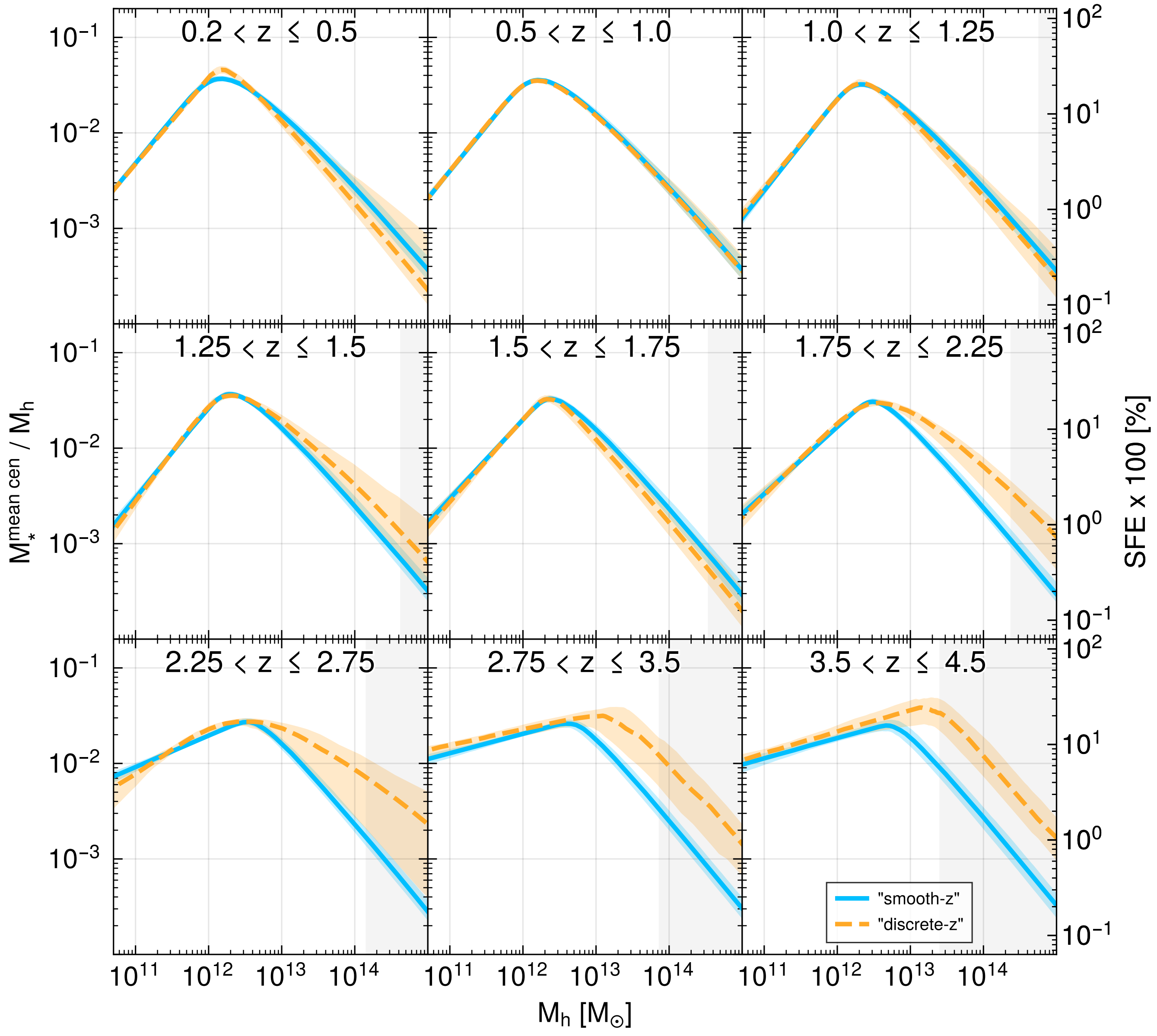

**Figure 15.** Mean stellar–halo mass ratio of central galaxies, $M_{*,\mathrm{meancen}}/M_h$ as a function of halo mass for all 9 $z$-bins based on "smooth-$z$" fits in blue compared with the "discrete-$z$" results in yellow. The SFE $(=(M_*^{\mathrm{mean\ cen}}/M_h)/f_b)$ in percentages is indicated on the rightmost ordinates. The gray shaded regions represent halo masses not directly probed in our analysis, calculated based on "smooth-$z$" fits as described in Section 5.3.

## Appendix C
## Redshift Evolution of the SMFs for Halos with Fixed Masses

We derive the conditional (on halo mass) SMFs for our best-fit models by adjusting the integration limits in Equation (18). The SMFs of 0.5 dex wide halo mass bins from $10^{11.5}$ to $10^{13.5}$ $M_\odot$ are shown in Figure 16, with the legend at the top indicating the four halo mass bins. The contributions from centrals and satellites are shown separately through dashed and solid curves, respectively, and the shaded regions represent the $1\sigma$ uncertainties. The central galaxies see a tight correlation with all halo mass bins out to $z = 4.5$; their SMFs peak at a certain stellar mass which increases with halo mass, and falls off abruptly on either end. This tight correlation of centrals with halo mass is not enforced by HOD modeling, as we fitted for the scatter in stellar mass at fixed halo mass, $\sigma_{\log M_*}$, which is linked to the scatter in $M_h$ at fixed $M_*$, $\sigma_{\log M_h}$, relevant here. As described in L11, since the SHMR behaves like a power law at low-mass, $\sigma_{\log M_h}{=}\beta{\times}\sigma_{\log M_*}$, where $\beta$ controls the low-mass slope. At high mass, $\sigma_{\log M_h}$ rises at a rate set by $\delta$ and $\gamma$. In our fits, $\sigma_{\log M_*}$ was allowed to reach 0.5 dex, i.e., $\sigma_{\log M_*}\in$ [0.001, 0.5], but it never surpassed $\sim$0.3 dex in any $z$-bin, indicating that the tight correlation of centrals with $M_h$ we observe was not enforced via HOD modeling.

On the other hand, satellites have large scatter; there is almost no preference for the halo mass for satellites less massive than $M_* \sim 10^{10} M_\odot$. This highlights how simple AM by rank ordering halos and galaxies with mass is not enough. Clustering provides additional information which can then be used to constrain the contribution of number densities to central and satellite galaxies individually. Satellites more massive than $M_* \gtrsim 10^{10} M_\odot$ are naturally hosted according to their mass; more massive satellites reside in more massive halos hierarchically.

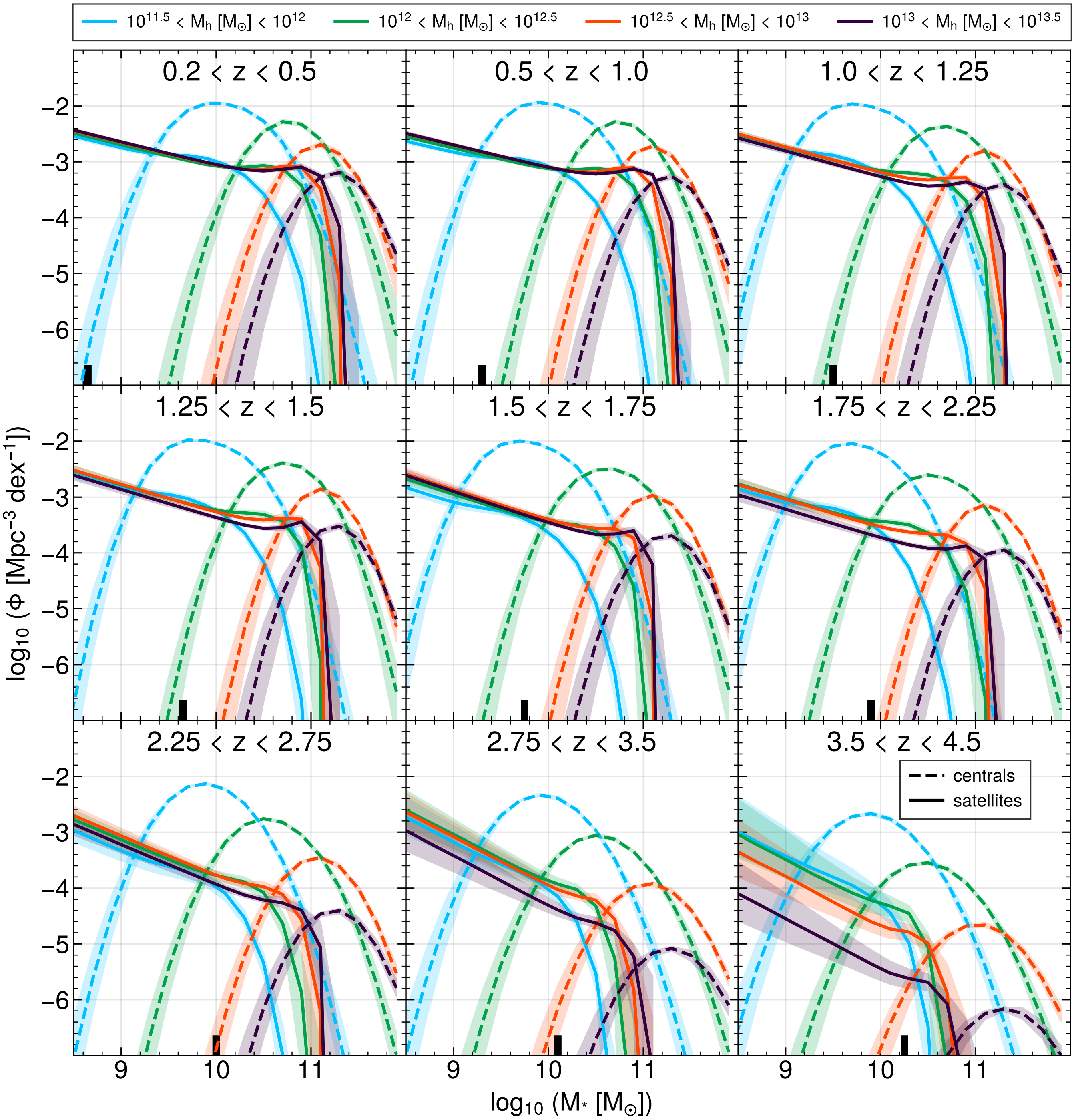

**Figure 16.** SMFs for centrals (dashed curves) and satellites (solid curves) within 0.5 dex wide bins of halo masses highlighted in different colors and indicated in the legend. The lowest stellar mass threshold considered within each $z$-bin, due to the stellar mass completeness (as described in Section 2.2 and quoted in Table 1), is indicated with a black marker at the bottom of each panel.

## Appendix D
## Redshift Evolution of the Satellite Fractions

As shown in Section 5.5, satellites start to contribute noticeably to the stellar mass budget of the halos beyond $M_h \sim 10^{12}\ M_\odot$. Their contributions continue to rise and end up dominating the stellar mass budget beyond $M_h \sim 10^{13}\ M_\odot$ in the group and cluster-scale halos. As satellites are acquired through mergers, analyzing the redshift evolution of satellite fraction, $f_{\rm sat}$ (calculated using Equation (9)), can shed light on the evolution of the environment of these halos.

In Figure 17, we show the stellar mass threshold dependence of $f_{\rm sat}$ in bins of redshift (left panel) as well as the redshift dependence of $f_{\rm sat}$ for different stellar mass thresholds (right panel). Within each $z$-bin, $f_{\rm sat}$ only weakly grows as the stellar mass threshold is lowered to include more and more galaxies. For instance, in the $1.0 < z < 1.25$ bin (shown in cyan), $f_{\rm sat}$ only evolves from $\sim$18% to $\sim$23%. Indeed at any redshift, lowering the stellar mass threshold does not increase the $f_{\rm sat}$ by more than $\sim$6% within the stellar mass ranges considered. Furthermore, even within the lowest $z$-bin with the lowest stellar mass threshold ($M_* > 10^{8.65} M_\odot$) considered here, where $f_{sat}$ is maximized as expected, it only reaches ∼28%.

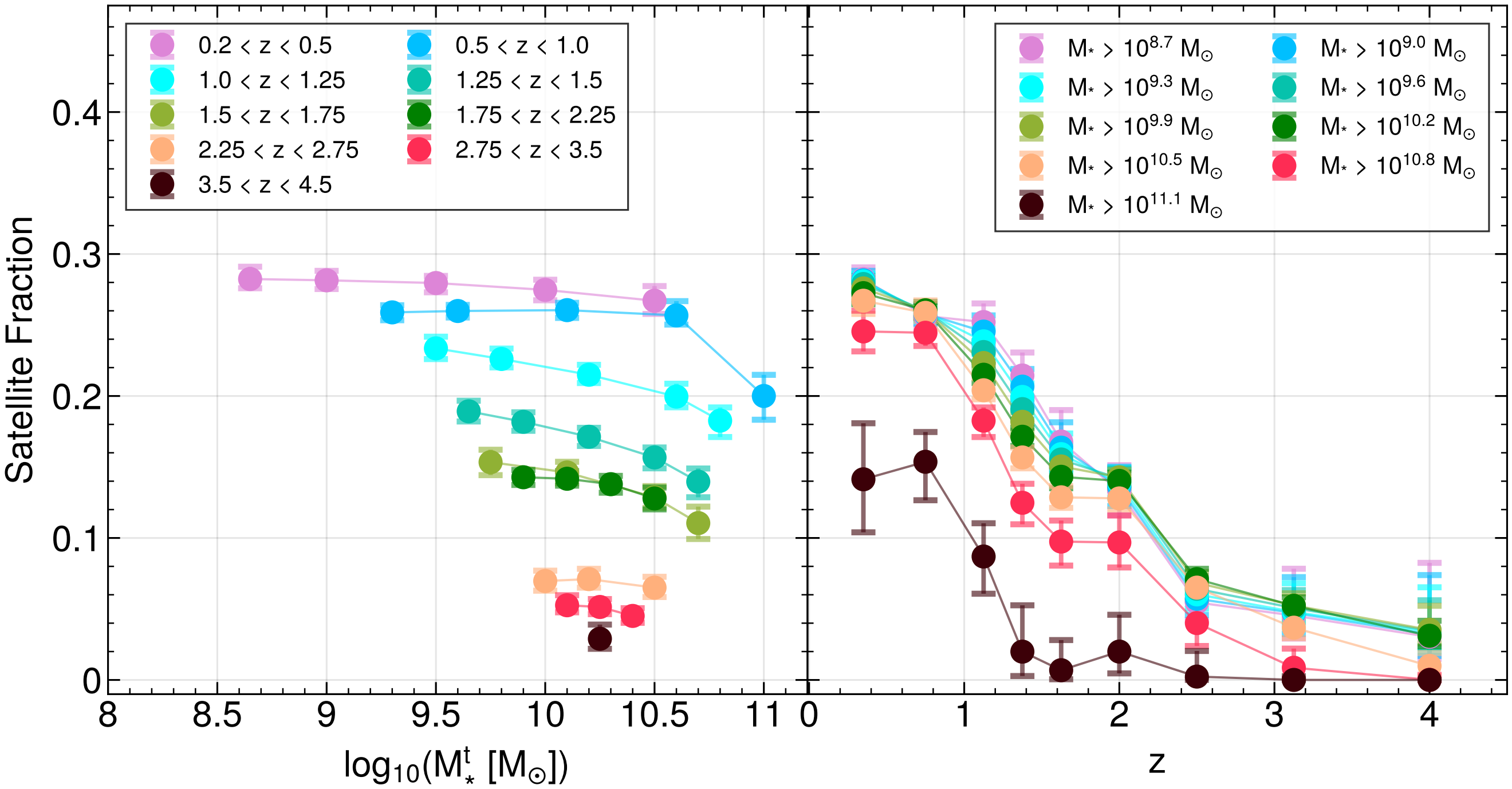

**Figure 17.** Left: satellite fractions within each $z$-bin (as indicated in the legend) as a function of the stellar mass threshold, $M_*^t$, of the subsamples. Right: satellite fractions of increasing stellar mass thresholds (as shown in the legend) as a function of redshift.

The plateauing of $f_{sat}$ with decreasing stellar mass threshold, visible at $z < 1$, can be attributed to the much higher abundance of the low-mass halos and the sharp decline at the high-mass end of the HMF. Therefore, as the stellar mass threshold is lowered to include more and more low-mass galaxies, they are more likely to be allocated as central galaxies in highly abundant low-mass halos than as satellites in rare high-mass halos. Even though the occupation number of satellites increases as a power law as halo mass increases, due to the shape of the HMF, there are just too few high-mass halos available to host many low-mass satellites. Therefore, low-mass galaxies are more often hosted as centrals in low-mass halos (M. Shuntov et al. 2022).

While the dependence of $f_{sat}$ on stellar mass threshold is relatively weak, it depends much more strongly on redshift such that it increases sharply as the Universe ages as seen by the vertical offsets between different $z$-bins again in the left panel of Figure 17. This is shown more clearly in the right panel of Figure 17, where $f_{sat}$ sees an overall increasing trend with decreasing redshift, and plateaus at $z \lesssim 0.75$. In general, at each redshift, the lower the stellar mass threshold, the higher its $f_{sat}$, unsurprisingly.

## ORCID iDs

Kumail Zaidi https://orcid.org/0000-0002-1163-7790
David A. Wake https://orcid.org/0000-0002-6047-1010
Danilo Marchesini https://orcid.org/0000-0001-9002-3502
Kartheik Iyer https://orcid.org/0000-0001-9298-3523
Adam Muzzin https://orcid.org/0000-0002-9330-9108
Casey Papovich https://orcid.org/0000-0001-7503-8482
Jacqueline Antwi-Danso https://orcid.org/0000-0002-0243-6575
Karl Glazebrook https://orcid.org/0000-0002-3254-9044
Ivo Labbé https://orcid.org/0000-0002-2057-5376